\documentclass[sigconf]{acmart}

\usepackage[english]{babel}
\usepackage[load-configurations={abbreviations,binary},binary-units]{siunitx}
\usepackage{blindtext}
\usepackage{subcaption}
\usepackage{caption}
\usepackage{algorithm}
\usepackage{algorithmic}
\usepackage{tikz}

\usepackage[medium, compact]{titlesec}
\titlespacing*{\section}{1pt}{3.5pt}{2pt}
\titlespacing*{\subsection}{1pt}{3pt}{1.5pt}
\titlespacing*{\subsubsection}{1pt}{3pt}{1.5pt}
\newcommand*\circled[1]{\tikz[baseline=(char.base)]{
            \node[shape=circle,draw,inner sep=2pt] (char) {#1};}}
\renewcommand\footnotetextcopyrightpermission[1]{} 
\setcopyright{none}

\setcopyright{none}
\acmDOI{}

\acmISBN{}

\acmPrice{}

\newcommand{\gz}[1]{\noindent{{\bf \fbox{GERD} {\it#1}}}}

\newcommand{\Fig}[1]{Figure~\ref{#1}}
\newcommand{\Sec}[1]{Section~\ref{#1}}

\newcommand{\secref}[1]{\S\ref{#1}}
\newcommand{\figref}[1]{Figure~\ref{#1}}

\newcommand{\ns}[1]{\SI{#1}{\nano\second}}
\newcommand{\us}[1]{\SI{#1}{\micro\second}}
\newcommand{\mis}[1]{\SI{#1}{\milli\second}}
\newcommand{\KiB}[1]{\SI{#1}{\kibi\byte}}
\newcommand{\MiB}[1]{\SI{#1}{\mebi\byte}}
\newcommand{\GiB}[1]{\SI{#1}{\gibi\byte}}
\DeclareSIUnit{\cycles}{cycles}

\DeclareSIUnit{\seconds}{secs}

\newcommand{\KB}[1]{\SI{#1}{\kilo\byte}}
\newcommand{\MB}[1]{\SI{#1}{\mega\byte}}
\newcommand{\GHz}[1]{\SI{#1}{\giga\hertz}}

\newcommand{\Gbps}[1]{\SI{#1}{\giga bps}}

\newcommand{\system}{PL2\xspace}
\newcommand{\rts}{reserve\xspace}

\newcommand{\cts}{grant\xspace}

\newcommand{\reth}{raw Ethernet\xspace}
\newcommand{\oreth}{raw Ethernet (optimal)\xspace}
\newcommand{\breth}{raw Ethernet\xspace}
\newcommand{\pleth}{PL2 over Ethernet\xspace}

\newcommand{\rsv}{RSV\xspace}
\newcommand{\grt}{GRT\xspace}

\hypersetup{draft}
\begin{document}
\title{\system: Towards Predictable Low Latency in Rack-Scale Networks }
\author{\LARGE Yanfang Le $^{\dagger}$, Radhika Niranjan Mysore $^{\dagger \dagger}$, Lalith Suresh $^{\dagger \dagger}$, Gerd Zellweger$^{\dagger \dagger}$, \\Sujata Banerjee$^{\dagger \dagger}$, Aditya Akella$^{\dagger}$, Michael Swift$^{\dagger}$ \\
University of Wisconsin-Madison$^{\dagger}$, VMware Research$^{\dagger \dagger}$}





\begin{abstract}
    High performance rack-scale offerings package disaggregated pools of compute, memory and storage hardware in a single rack to run diverse workloads with varying requirements, including applications that need low and predictable latency.  The intra-rack network is typically high speed Ethernet, which can suffer from congestion leading to packet drops and may not satisfy the stringent tail latency requirements for some workloads (including remote memory/storage accesses).
In this paper, we design a Predictable Low Latency (\system)  network
architecture for rack-scale systems with Ethernet as interconnecting fabric. \system leverages programmable Ethernet switches to carefully schedule packets such that they incur no loss with NIC and switch queues maintained at small, near-zero levels.  
In our \Gbps{100} rack-prototype, \system keeps 99th-percentile memcached RPC latencies under \us{60} even when the RPCs compete with extreme offered-loads of 400\%, without losing traffic. Network transfers for a machine learning training task complete 30\% faster than a receiver-driven scheme implementation modelled after Homa (222ms vs 321ms 99\%ile latency per iteration). 

\end{abstract}

\maketitle

\section{Introduction}
\label{sec:intro}

Rack-scale data center solutions like Dell-EMC VxRail~\cite{vxrail} and Intel RSD~\cite{intelrsdwhitepaper} have emerged as a new building block for modern enterprise, cloud, and edge infrastructure. These rack-units have three key characteristics; First is the increasing use of \textit{resource disaggregation and hardware accelerators} within these rack-units like GPUs and FPGAs~\cite{intelrsdwhitepaper}, in addition to high-density compute and storage units. Second, \textit{Ethernet} is by far the dominant interconnect of choice within such racks, even for communication between compute units, storage units and accelerators (e.g., Ethernet-pooled FPGA and NVMe in Intel RDS~\cite{intel-arch}). Third, these racks are deployed in a wide range of enterprise and cloud customer environments, running a heterogeneous mix of modern (e.g., machine learning, graph processing) and legacy applications (e.g., monolithic web applications), making it \textit{impractical to anticipate traffic and workload patterns}.

Rack-scale networks\footnote{The network extending between NICs of such rack-units across the top-of-rack (ToR) switch.} need to satisfy the key requirements of \textbf{uniform low latency} and \textbf{high utilization}, irrespective of where applications reside, and which accelerators they access (e.g., FPGA vs. CPU vs. GPU). 
However, three key obstacles stand in the way of achieving these goals because of the above-mentioned characteristics.
First, the rack-scale network \textbf{must be transport-agnostic}, a necessity in environments with \emph{(a)} heterogeneous accelerator devices that have different characteristics\footnote{For example, FPGA stacks will not be connection-oriented due to scaling issues~\cite{mothyfpga} and GPUs will not have a single receiver stack~\cite{gpunet}).} than CPU network stacks~\cite{gpunet, flexnic, fpgaof}, and \emph{(b)} increasing use of CPU-bypass networking~\cite{mtcp, erpc,rocev2}. 
Second, Ethernet is not a lossless fabric, and yet, our experiments (\secref{sec:motiv}) on a 100G testbed confirm that \textbf{drops, not queueing}, are the largest contributor to tail latency pathologies. Third, the design must be \textbf{workload-oblivious} -- given that we cannot anticipate traffic and workload patterns across a broad range of customer environments, it is impractical to rely on state-of-the-art proposals (\S\ref{sec:related}) that hinge on configuring rate limits or priorities using a-priori knowledge of the workload.

In this paper, we present {\em Predictable Low Latency} or \system, a rack-scale lossless network architecture that achieves
low latency and high throughput in a transport-agnostic and workload-oblivious manner. \system reduces NIC-to-NIC latencies by proactively avoiding losses.
\system supports a variety of message transport protocols and gracefully accommodates increasing numbers of flows, even at 100G line rates. It neither requires a-priori knowledge of workload characteristics nor depends on rate-limits or traffic priorities to be set based on workload characteristics. 

To achieve these goals, senders in \system explicitly request a switch buffer reservation for a given number of packets, a \emph{packet burst}, and receive notification as to when that burst can be transmitted without facing any cross traffic from other senders. \system achieves this form of centralized scheduling even at 100G line rates by overcoming the key challenge of carefully partitioning the scheduling responsibility between hosts in the rack and the Top-of-Rack (ToR) switch. In particular, the end-host protocol is kept simple enough to accommodate accelerator devices and implementations within NICs (\S\ref{subsec:low_load}), whereas the timeslot allocation itself is performed in the ToR switch at line rate (as opposed to doing so on a host, which is prone to software overheads).

In summary, our contributions are:
\begin{itemize}
\item The \system design that embodies novel yet simple algorithms for lossless transmissions and near-zero queuing within a rack
\item A \system implementation using a P4 programmable switch and an end-host stack that leverages Mellanox's state-of-the-art VMA message acceleration library~\cite{mellanox-vma}
\item A comprehensive \system evaluation on a \Gbps{100} prototype, supporting three different transports (TCP, UDP and Raw Ethernet), all benefiting from near-zero queueing in the network. Compared to VMA, we demonstrate up to 2.2x improvement in the 99th percentile latency for the Memcached application; a 20\% improvement to run a VGG16 machine learning workload; and near-optimal latency and throughput in experiments using trace-based workload generators.
\end{itemize}

\section{Motivation}
\label{sec:motiv}
The primary goal of \system is to provide uniformly low-latency across Ethernet rack-fabrics, while achieving high-utilization.
We take inspiration from prior work around low and predictable latency within data center networks~\cite{homa,pfabric,hull,D2TCP,neverthanlate,ndp,phost,expresspass,dctcp,timely,dcqcn,vegas,acdc,cubic,hpcc,fastpass,flowtune,picnic,silo,eyeQ,predicatble-datacenter,qjump}, but find that rack-scale networks provide a rich set of new challenges.

\subsection*{Rack-scale characteristics and implications}

\paragraph{\textbf{1. Even as line-rates increase, intra-rack RTTs are not getting smaller.}}~\cite{homa} measured \us{5} end-to-end RTT on a \Gbps{10} testbed with a single ToR switch, inclusive of software delays on the servers. A 64B packet still has an RTT of \us{5} in our \Gbps{100} rack-scale \system prototype. Even though the network transmission times reduce proportionally with increasing line-rates, switching-delays, NIC hardware delays, and DMA transfer delays at end-hosts have remained about the same, and these delays together dominate transmission delays in a rack. Forward-error-correction delays increase as line-rates increase, and can add variability of up to 1-\us{2} in a \Gbps{100} network. This implies that rack-scale networks are able to transfer more data in the same RTT as interconnect speeds increase. Flows up to \KB{61} can complete within 1 RTT with a \Gbps{100} backplane as opposed to \KB{6} for a \Gbps{10} backplane. For congestion control protocols and predictable latency schemes to be effective for flows below these sizes, they will need to converge in sub-RTT timeframes.


\paragraph{\textbf{2. Even as rack-densities go up, network buffering in ToR switches is not getting bigger.}} Shallow buffers are even more critical to a disaggregated rack, because buffering adds latencies to network transfers. However, the implication of this trend is that microbursts can over-run shared output switch-buffers and cause drops. For instance, a \MB{2} buffer in a ToR switch with \Gbps{100} ports provides buffering for just \us{160} which means only 8 simultaneous transfers of 2Mbits can be sustained 
before the switch starts dropping packets. Today's rack-scale networks support up to 64  rack-units~\cite{vxrail-datasheet}, where each end-system can have tens of thousands of ongoing transfers. 
At that scale a 2 MB can be overrun with only 6 simultaneous \us{5} (1 RTT) network transfers per rack-scale unit. 
In short, as rack-densities go up, drops due to microbursts can be frequent. Therefore, assumptions made by congestion protocols like~\cite{homa,ndp} that the network-core (ToR switch in the case of racks) is lossless, no longer holds.

\paragraph{\textbf{3. Rack-scale traffic is ON/OFF traffic~\cite{benson}}}
We believe this trend will continue with network traffic generated by accelerators. Measuring traffic-demands in such environments is hard, let alone learning about workload-characteristics; traffic demands at ns-timescales will be different compared to $\mu$s timescales and ms-timescales~\cite{bursts}. Workload churn and different application mixes  adds to the unpredictability. 

Coming up with effective rate-limits~\cite{picnic, silo, eyeQ, predicatble-datacenter, qjump}, and readjusting these rate-limits with changing traffic-conditions in time (i.e., less than an RTT) is impossible; so is setting relative packet priorities~\cite{homa} effectively~\cite{bwe} so that important traffic is never lost or delayed.  In our experience neither rate-limits nor priority prescription is an answer to congestion-control within a rack.


\subsection*{Drops cause the most noticeable tails}

Based on the above three observations, we believe that new techniques are required to ensure low-latency and high-utilization within a rack-scale network. We hinge the \system design on the observation that drops, not queuing cause the most noticeable tails.
\begin{figure*}[!t]
      \begin{subfigure}{0.33\textwidth}
     \centering
	   \includegraphics[width=.95\linewidth,height=0.45\textwidth]{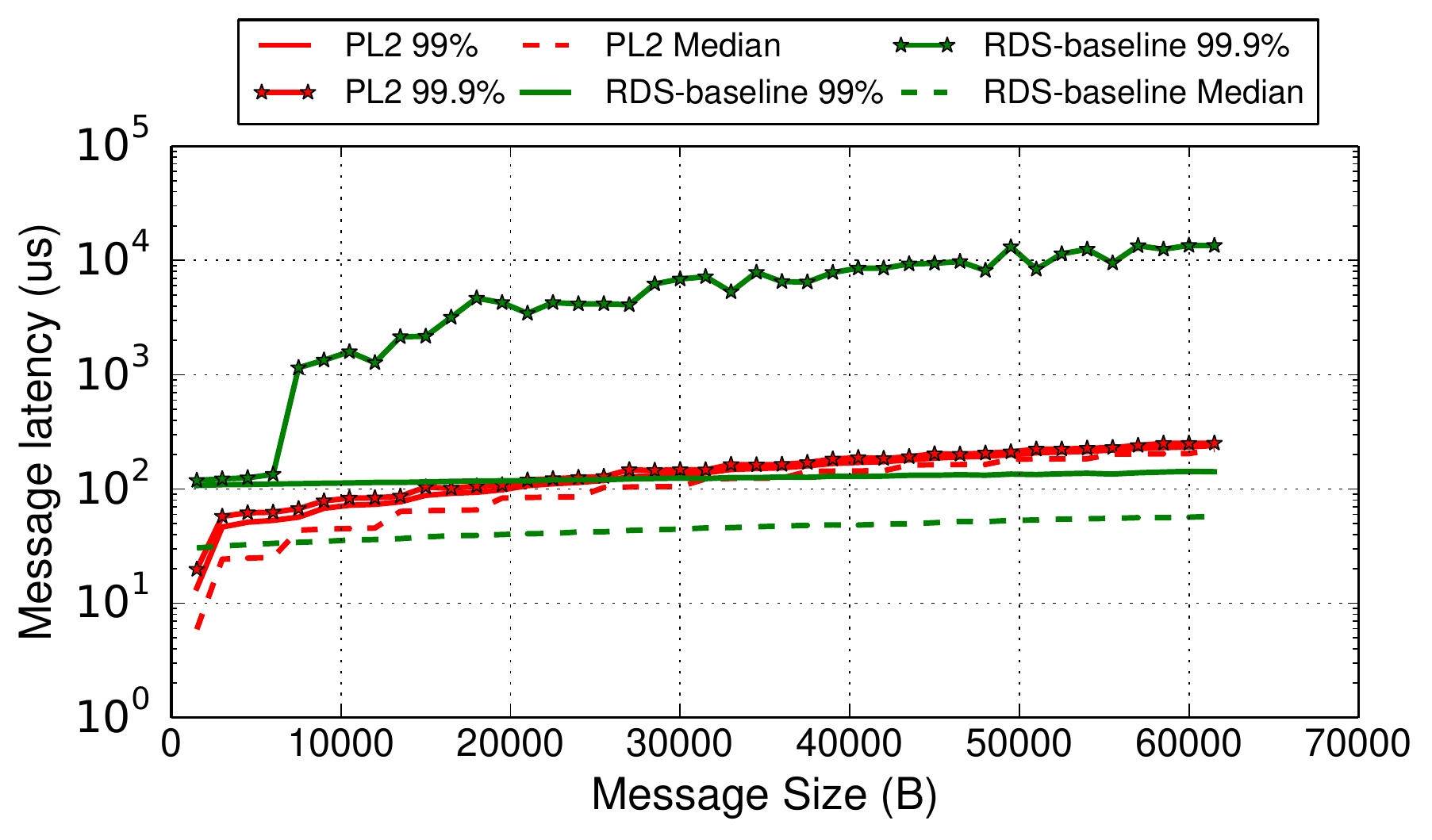}
	   \caption{\label{fig:microburst-mesg-lat} Message latency}
	   \label{fig:microbursts-message-latency}
   \end{subfigure}
     \begin{subfigure}{0.33\textwidth}
     \centering
	   \includegraphics[width=.95\linewidth]{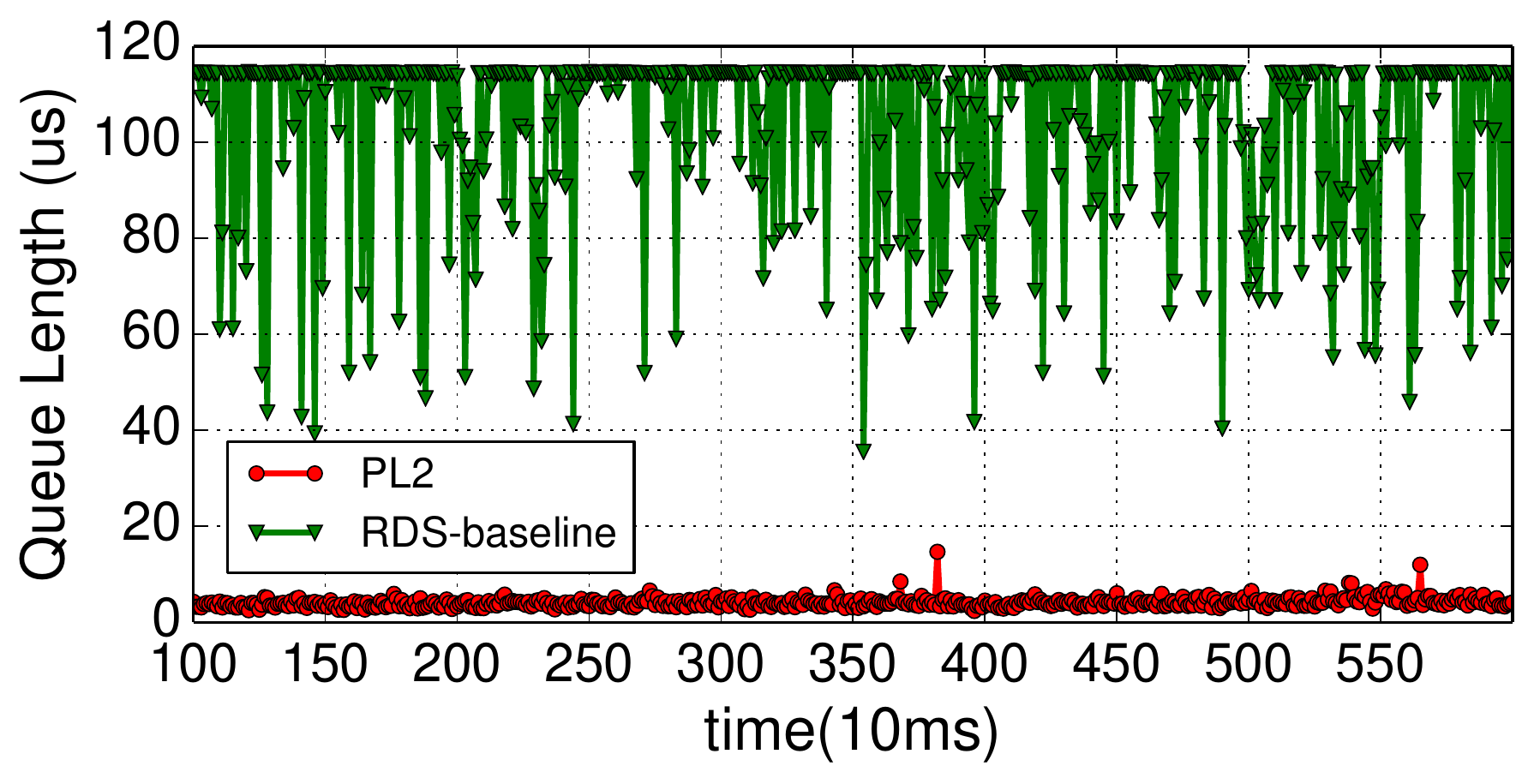}
	   \caption{Switch queuing delay}
	   \label{fig:microbursts-queue-length} 
   \end{subfigure}
         \begin{subfigure}{0.33\textwidth}
     \centering
	   \includegraphics[width=.95\linewidth]{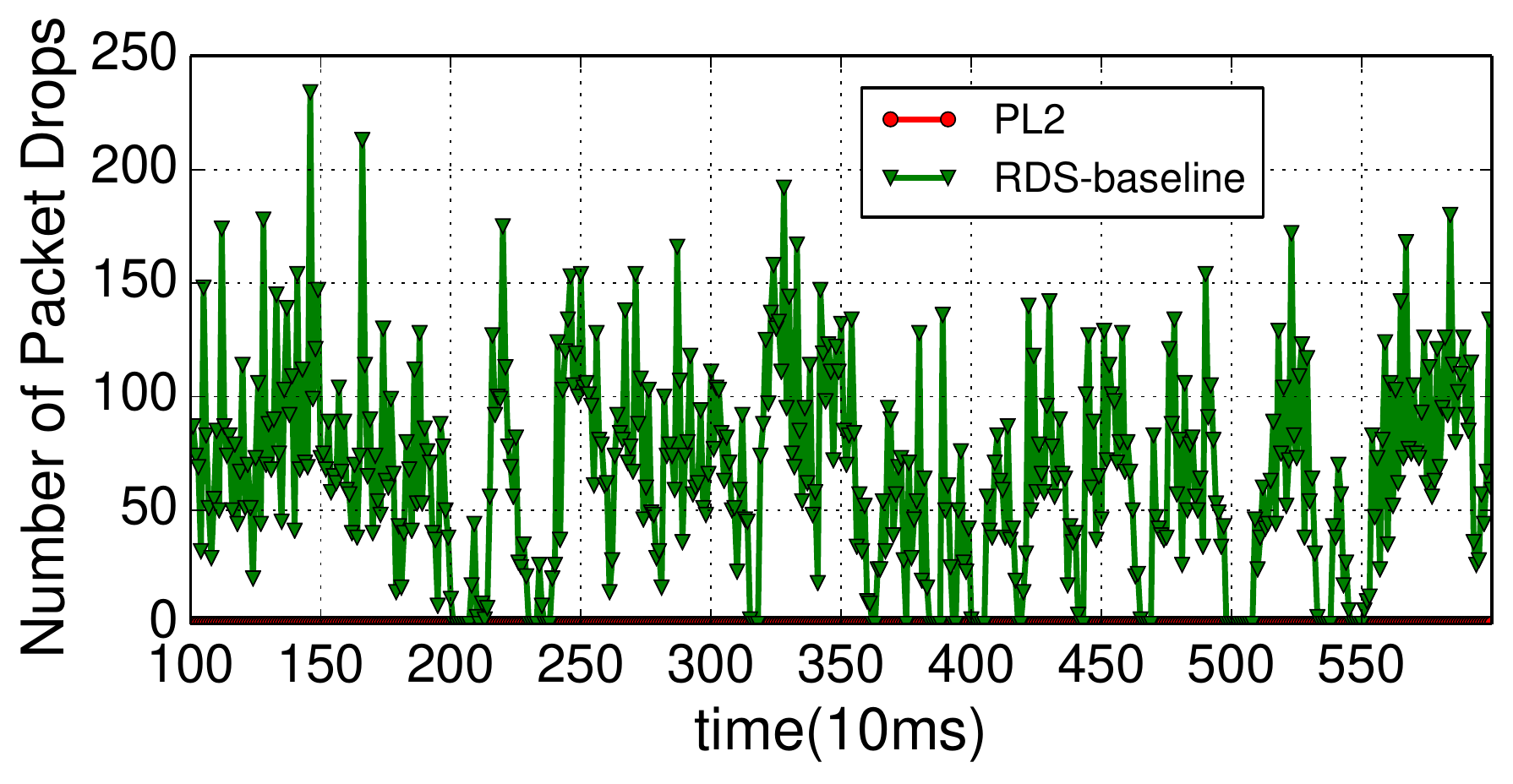}
	   \caption{\label{fig:microburst-packets-drop} Packets drop per $10$ms}
	   \label{fig:microbursts-packets-drop}
   \end{subfigure}
   \caption{Messages latency, switch queuing delay and packets drop during microburst.}
   \label{fig:microbursts}
   \vspace{-10pt}
\end{figure*}

We illustrate this point with an experiment that introduces microbursts in our \Gbps{100} testbed even when the network is partially loaded, by using 5-way incast of roughly \Gbps{18} traffic per sender. All messages transmitted in our experiment can be transmitted within a single RTT. As shown in \Fig{fig:microbursts-message-latency}, the 99\%ile latencies experienced by a receiver-driven scheme (RDS-baseline) based on Homa (described in detail in \Sec{sec:eval}) correspond to the maximum output-buffering available in the ToR switch (around \us{120} in \Fig{fig:microbursts-queue-length}), while the 99.9\%ile latencies correspond to delays due to drops, and are two orders of magnitude higher. Reducing the drop-timeout in our implementation only increases drops, while only slightly reducing the 99.9\%ile latencies.

In contrast, \system's 99\%ile and 99.9\%ile latencies are similar to its median latencies, and it does not experience drops. \system is not impacted by microbursts; it keeps buffer occupancy low (maximum of \KiB{200}). \Fig{fig:microbursts-packets-drop} shows the drops (around 0.1\%) experienced by RDS-baseline over time.

\section{Related Work}
\label{sec:related}
\paragraph{\textbf{Priority Flow Control (PFC)~\cite{pfc}}} PFC is a closely related L2 mechanism that can also counter loss in a rack-scale network. Configuring PFC for correct operation is notoriously hard, even at rack-scale (see PXE booting issues in~\cite{roce-scale}), and turning on PFC in a rack that is part of a larger data center can be disruptive. Even within a single rack, PFC's coarse-grained mechanisms of providing losslessness across less than 8 traffic classes requires operators to choose between high utilization and lossless behavior because congestion in one downstream class can result in multiple unrelated senders receiving PAUSE frames due to HOL blocking~\cite{dtc}.

\paragraph{\textbf{Rack-scale interconnects}}
Several custom designs for rack scale interconnects have been proposed;~\cite{r2c2} propose direct-connect topologies, and~\cite{shoal} proposes circuit switched connectivity within a rack.~\cite{genz, omnipath,ccix} propose cache-coherent interconnects at rack scale to enable new computation models at rack-scale. \system differs fundamentally from all of these in vision. Even though other designs might perform better, \system's goal is 
to allow traditional and commercially available rack-scale architectures to continue to avail the cost and operational ease benefits of Ethernet interconnects, but also to get predictable latency benefits; 

\paragraph{\textbf{Predictable latency}}
\cite{picnic, silo, eyeQ, predicatble-datacenter, qjump} provide predictable latency by resource isolation among tenants or applications of a data center. All of these use rate-limit based network admission control to ensure isolation, and require a-priori knowledge of traffic characteristics (application mixes, demands). They typically dynamically readjust rate-limits based on new demand, but require considerable time to do so. For example,~\cite{picnic} requires a few RTTs for convergence. Often these systems can rely on inputs from applications, tenant requirements or systems like Bwe~\cite{bwe} to determine rate-limits. As described in \Sec{sec:motiv}, \system cannot leverage these ideas in the rack-scale context. 

\paragraph{\textbf{Congestion-control}}
Our work follows the long history of low-latency, high-throughput congestion-control mechanisms. 
\cite{homa,ndp,phost,expresspass} propose software-based receiver-side scheduling targeted towards \Gbps{10} data center networks. Some of these schemes~\cite{homa,ndp} rely on the assumption that the network core does not experience loss; an assumption that is invalid in the rack-scale context (\Sec{sec:motiv}).

Recent proposals suggest that starting new flows at line-rates~\cite{ndp, homa}, or over-committing downlinks~\cite{homa} could speed up network transfers; the experiment described in \Fig{fig:microbursts} verifies that this idea trades off tail-latency for improved minimum and median latencies, which may not be beneficial~\cite{tail-at-scale}.
 
Homa, pFabric, and others~\cite{homa, pfabric, hull, D2TCP, neverthanlate, qjump} depend on prior knowledge of application traffic characteristics for providing benefits over other schemes; something that may be difficult due to shifting or short-lived workloads~\cite{google-trace}. 

Most congestion control schemes proposed  
~\cite{dctcp, timely, dcqcn, vegas, acdc, cubic, hpcc} rely on layer $3$ and above to remedy congestion after it is observed either by way of delay, ECN marks, buffer occupancy or packet loss~\cite{cp}. They require at least an RTT to respond to congestion and are too slow to prevent drops in a rack-scale network.

Fastpass~\cite{fastpass} and Flowtune~\cite{flowtune} proactively request permission to send packets and are the closest prior schemes to \system. Fastpass decides when to send a burst of packets while Flowtune decides the rate to send a burst of packets. Their centralized arbiters, however, are host-based and cannot keep up with \Gbps{100} line rates because they are bottlenecked both by scheduling software latencies and the downlink to the arbiter. 
Control packets to the centralized arbiter can also be dropped arbitrarily depending on NIC polling rates; these issues make centralized scheduling at an end host too fallible to be efficient and effective at \Gbps{100}.

Most of these schemes (except for Fastpass and Flowtune) do not tackle the problem of being transport-agnostic; they rely on all traffic using the same end-host-based congestion control\footnote{This is true of Homa also, which reorders traffic based on message size and would not interact well with TCP.}. These schemes do not interact well with traffic that does not have congestion control. In \system, the rack-scale interconnect intercepts traffic from all higher layers, and is therefore
well suited to offer the properties we seek.


\section{\system Design}
\label{sec:design}
\system transforms the rack-scale network into a high-performance lossless low-latency interconnect. 
At the heart of \system is an algorithm for scheduling packet bursts at line rate using the
\emph{switch dataplane}, where a packet burst is simply a bounded number of Ethernet frames.


\system is designed to be losslessness via proactive congestion control, while at the same time, being both transport-agnostic and, workload-oblivious.

We achieve losslessness via the scheduling algorithm, which implements a timeslot
reservation scheme where each sender transmits only at its assigned timeslot, reducing cross-traffic. Since the scheduling function is placed in the switch dataplane in the network layer, it can schedule for packet bursts corresponding to all transports. The switch is one hop away from all hosts; therefore hosts can access the scheduling function at less than end-to-end RTT; we further eliminate scheduling overheads where possible. \system does not assume a-priori knowledge of traffic patterns or workloads.

In the following sections, we explain our scheduling algorithm first, followed by practical 
hardware constraints that inform its design.
It is worth mentioning, for readers familiar with Fastpass, that we are unable to borrow the timeslot allocation scheme in Fastpass due to these hardware constraints; \system switch scheduling algorithm trades off optimal scheduling for speed. We elaborate on this towards the end of the next subsection.

\subsection{\system scheduling algorithm}

\paragraph{Timeslot reservation overview}

\begin{figure}[t]
 \vspace{-15pt}
\begin{center}
\includegraphics[width=0.7\textwidth,trim=50 250 0 0]{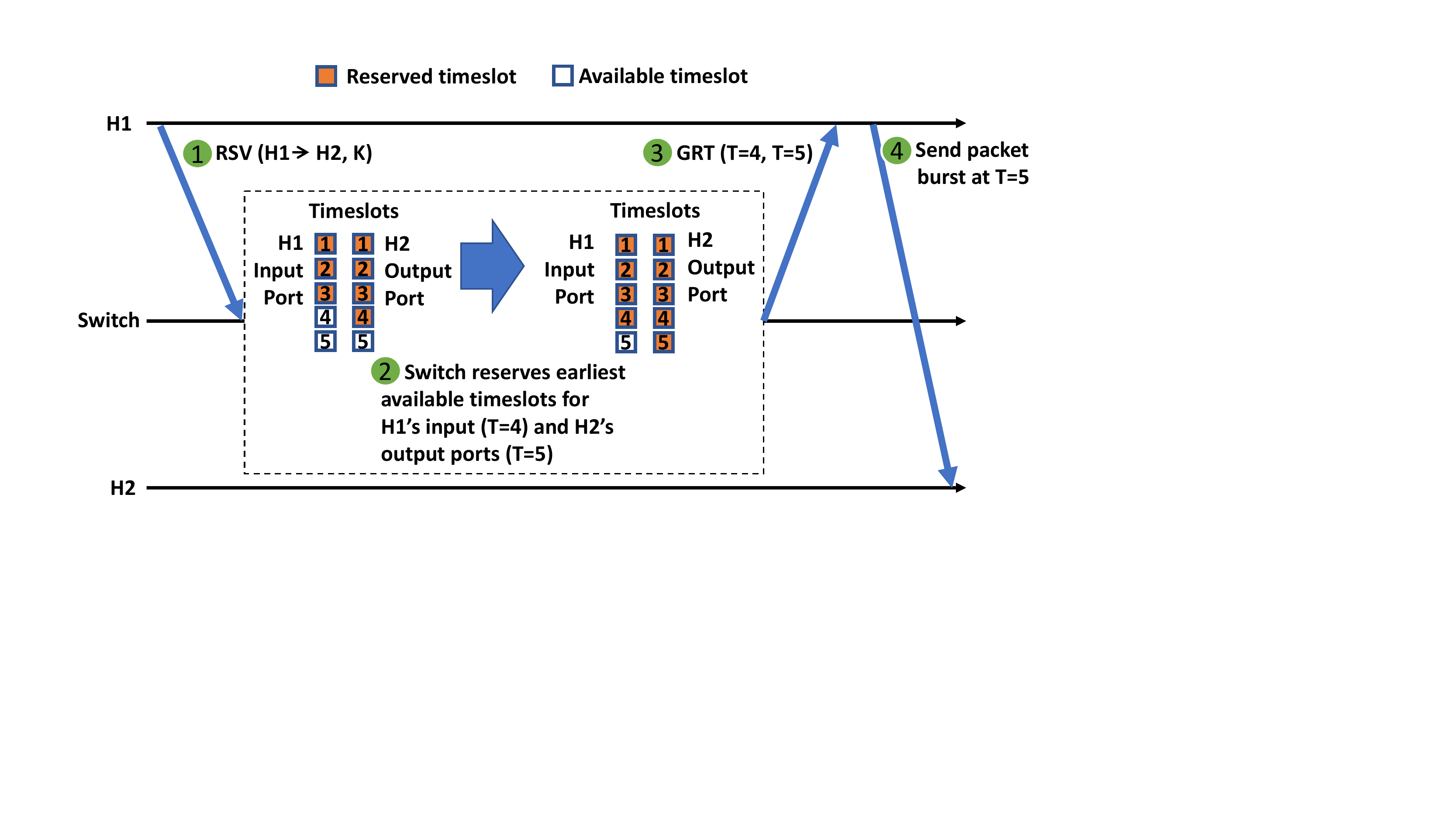}
\end{center}
 \caption{\small{\label{fig:arch} Scheduling example in \system, with host H1 sending a packet
 burst to H2. \protect\circled{1} H1 sends an \textit{\rsv} to the switch to make a reservation. 
 \protect\circled{2} The switch maintains timeslot reservations for the input and output ports
 connected to every hosts. It reserves the earliest available timeslots
 on H1's input port ($T=4$) and H2's
output port ($T=5$). \protect\circled{3} The switch notifies H1 of these timeslots
through a \textit{\grt} message. \protect\circled{4} To avoid queuing, H1 then
transmits at the maximum of the two timeslots indicated in the \grt, which is
$T=5$.}}\vspace{-0.1in} 
\end{figure}

Conceptually, our scheme maintains a list of \emph{timeslots} per input and
output buffer for each port. In our current implementation, we define a
timeslot to be the time it takes to transmit an MTU sized packet out of a
buffer\footnote{ The ideal duration for a timeslot is system and workload
dependent. It should be chosen as a function of common (or minimum) message sizes
and link speeds to ensure high utilization; transmitting small messages in
large timeslots is wasteful. It's choice is also determined by the transmission granularity in a system; for e.g, a timeslot
that is in picoseconds is useless because current hardware cannot
transmit at such a fine granularity.}. To transmit a packet burst from host
$h$ to $h'$, we seek to reserve a timeslot $t$ on the switch input port
corresponding to $h$, and a timeslot $t'$ on the output port corresponding to
$h'$. Host $h$ then transmits at a `chosen timeslot' which is the greater of
timeslots $t$ and $t'$ to avoid a collision\footnote{Our design choice to always choose the greater of timeslots $t$ and $t'$ can create gaps in time when no packet is scheduled.}. The astute reader will observe that we could instead let the switch choose the transmission time in a centralized manner rather than pairwise, but hardware constraints prevent us from doing so (\secref{sec:design-hardware}).

Note, with the hosts choosing the transmission times, there is a risk of collision. With perfect scheduling, we would have no collisions, and need near zero buffering at the switch. However, \system uses a small amount of buffer space (less than 200KB in our \Gbps{100} testbed) to accommodate occasional collisions.




To run the above-mentioned scheme at line rate and within the constraints of
switching hardware (outlined in \S\ref{switch_logic}), we designed an algorithm with the following
division of logic between the switch and the hosts: \emph{(i)} to
transmit a packet burst of size K\footnote{$K$ is technically the number of timeslots a sender is allowed to reserve at a time.}, 
hosts send {\emph \rts} or \rsv packets to the switch to
reserve timeslots, \emph{(ii)} the switch grants a reservation of size K and
responds to the host with a {\emph \cts} or \grt packet, which specifies the earliest available timeslot for the
corresponding input port and output port, \emph{(iii)} the host then picks the maximum of the two timeslots to avoid a
collision, \emph{(iv)} finally, the host converts the timeslot into a waiting
time after which it transmits the packet burst. In doing so, the timeslot
reservation logic is divided between the switch and the host. Importantly,
both the switch and host logic stays simple and in line with switching
hardware constraints, which we describe below.
{
\small
\definecolor{gray2}{rgb}{0.8,0.8,0.8}
\begin{algorithm}[hbt!]
    \caption{Switch Scheduling Algorithm. Each highlighted block is a single P4 operation.}
    \begin{algorithmic}[1]
        \STATE \textbf{INIT:}
        \STATE $inReservation[port\ 1..port\ n] \gets \{0\}$
        \STATE $outReservation[port\ 1..port\ n] \gets \{0\}$ 
        \vspace{4mm}
        \STATE \textbf{INPUT:} packet
        \STATE $src \gets $source port of \rsv 
        \STATE $dst \gets $destination port requested 
        \IF {packet is a \rsv}
        \STATE \colorbox{gray2}{\strut packet.sendTimeslot $\gets$ inReservation[src]}\vspace{-0.1in}
        \STATE \colorbox{gray2}{\strut inReservation[src] += packet.demand}\vspace{0.05in}
        \STATE \colorbox{gray2}{\strut packet.recvTimeslot $\gets$ outReservation[dst]}\vspace{-0.1in}        
        \STATE \colorbox{gray2}{\strut outReservation[dst] += packet.demand}\vspace{0.05in}
        \STATE \colorbox{gray2}{send \grt}
        \ELSE 
		\STATE \colorbox{gray2}{$outReservation[dst]$ -= $1$}
        \STATE \colorbox{gray2} {$inReservation[src]$ -= $1$}
        \ENDIF
    \end{algorithmic}
    \label{algo:scheduler-switch}
    \vspace{-2pt}
\end{algorithm}
}
\subsubsection{Switch logic}
\label{switch_logic}

Algorithm~\ref{algo:scheduler-switch} shows the scheduling logic at the
switch. To stay ahead of packet transmissions, despite the delay of each
\rsv-\grt exchange, the switch creates a schedule of input and output
reservations for every port, in terms of timeslots. Each highlighted
gray box represents logic that can be implemented with a single P4 operation. 

At switch start up, the input and output reservations are initialized to zero
(lines 2-3). In response to \rsv packets, the switch sends back the next available
input and output timeslots for the requested transmissions (lines 8,10,12).
It also reserves enough timeslots for each \rsv request (lines 9,11). In
\figref{fig:arch}, the switch has reserved timeslot 4 at the source
port (connected to Host 1) and timeslot 5 at the destination port (connected to
Host 2) for transmission. Note that the reservation timeslots have not lined
up exactly at the two ports. We describe how the sending host uses these
timeslots next. Lines 13-15 describe switch logic for regular packets, in which the timeslot reservations for the packet are removed. 


\subsection{Hardware constraints}
\label{sec:design-hardware}
There are two key hardware constraints that Algorithm~\ref{algo:scheduler-switch} satisfies.
First, currently available programmable switching hardware cannot access more than one
stateful memory object (SMO) at a time in an operation, per packet, at line-rates. This is why \texttt{inReservation} and \texttt{outReservation} timeslot counters in Algorithm~\ref{algo:scheduler-switch} are accessed and updated in two separate operations.
Second, all pipelined network hardware can only read/modify/write SMOs once
during the processing of packets. If a SMO is updated or accessed twice during the same pipeline, it results in race-conditions across packets. This is why \texttt{inReservation} and \texttt{outReservation} have to be read and modified in a single atomic operation in Algorithm~\ref{algo:scheduler-switch}. 

In the absence of the first limitation, the switch could update both \texttt{inReservation} and \texttt{outReservation} to max(\texttt{in\-Reservation}, \texttt{outReservation}). If  it could also cache the computed timeslot in packet metadata, it could send this timeslot information in response to a \rsv message.
Since the chosen timeslot increases monotonically, this scheme removes any possibility of collisions.
In the current implementation, 
the switch relays 
\texttt{inReservation} and \texttt{outReservation} to the host, which computes the chosen timeslot to send at.

\paragraph{Why not implement Fastpass in a switch instead?}
Fastpass~\cite{fastpass} is able to perform timeslot allocation with maximal matching because the scheduler processes a list of all demands in the network at once; implementing such a scheme is impossible in the switch dataplane at line-rate, because switch pipelines cannot compare multiple \rsv packets in flight. 

In addition, Fastpass performs timeslot allocation using a bitmap table, that has a sender and receiver bitmap to track multiple timeslots. Allocation requires a bitwise \texttt{AND} of the sender and receiver bitmap, followed by a `set' on the first available timeslot. Supporting such an algorithm would require hardware to be able to access and set at least two stateful memory objects in a single operation. Maintaining a sliding window of timeslots is similarly hard to achieve with dataplanes today because they expose minimal timing API that are restricted to timestamping --- converting these timestamps to sliding windows requires accessing multiple stateful memory objects, one that maintains timestamp information, and another that maintains timeslot information, and updating them at line-rates for every \rsv packet.


\subsubsection{Host logic} 
\label{sec:hostlogic}
On the host side, for lossless transmission, senders must ensure that their
packet bursts do not collide with other transmissions at \textit{both} the
corresponding input and output ports on the switch. In fact, input and output 
ports at a switch might be shared by multiple hosts (like when a 100G link 
is divided into four 25G links and connected to four different hosts).
End hosts in \system therefore conservatively choose a timeslot to transmit
that is available on both the relevant switch input and output ports using the equation
$chosenTimeslot \gets max(sendTimeslot, recvTimeslot).$

The sender then transmits packets at the
\texttt{chosenTimeslot} by waiting for a period \texttt{waitingTime} calculated using the equation 
\begin{equation}
 \begin{split}
    waitingTime  & = timeslotWait - rsvGrtDelay \\
    & - pendingDataDelay,\\
      \end{split}
    \label{eq:waittime}
\end{equation}
where $timeslotWait = chosenTimeslot * MTU/linerate$
and \texttt{pendingDataDelay = bytes/linerate}.

The explanation for this calculation is as follows: The timeslot chosen for
transmission is \texttt{timeslotWait} seconds into
the future from when the reservation is made at the switch. This reservation
is conveyed back to the sender after \texttt{rsvGrtDelay/2} seconds. The
first packet of the transmission would again take approximately \texttt{rsvGrtDelay/2} seconds to reach the switch, after transmitting all the previously scheduled
packets at the sender NIC  (which would take \texttt{pendingDataDelay} seconds)\footnote{We ensure that \rsv-\grt packets do not wait behind data packets (if any) by prioritizing them using 801.1q}. The sender  therefore waits only for the remaining fraction of time to ensure that the first packet of the burst arrives in time at the switch. 

The aforementioned logic has some leeway with regards to where on the
host it runs. In our current userspace stack implementation, one instance 
of the \system host logic runs per application thread and each thread maintains at
most one outstanding \rsv-\grt exchange. We believe a better
implementation would be one where the host logic runs inside a NIC.
In that case, the NIC can allow sending threads to have one outstanding \rsv-\grt exchange per destination mac address, so that a thread sending messages simultaneously to multiple destinations can do so without encountering head-of-line blocking for messages to unrelated destinations.

\subsection{Setting the packet-burst size, $K$}
\label{sec:design-k}

A key parameter in \system is the packet-burst size, $K$. \system ensures stable queuing by proactively
scheduling transmissions such that when these packet bursts are transmitted, they encounter almost no queuing.
However, the timeslot reservations for the relevant input and output ports for a transmission determine
when packet bursts are transmitted. The switch reserves as many timeslots as needed for a transmission
based on the demand from the host, which is capped by $K$.

A small value of $K$ limits the amount of head-of-line blocking a burst introduces at its input and output ports and ensures that \system supports a large number of concurrent transmissions at any point in time.
However, when $K$ is too small, the overhead of the \rsv-\grt exchange dominates, lowering throughput and effective utilization.  Similarly, large values of $K$ help amortize the cost of the \rsv-\grt exchange delay, but increase head-of-line blocking because that causes the switch to reserve a burst of timeslots for the same host, potentially starving other hosts.

We find that $K=4$ works best for our 100G testbed prototype and in our simulation for all the workloads we tested, based on a parameter sweep. 
\secref{sec:impl} details the configurations in our testbed and simulations.

\subsection{Reducing scheduling overheads}
\label{subsec:low_load}
Each \rsv-\grt exchange enables senders to determine the waiting time before transmitting on a given input/output port pair. 
Under light loads, the waiting times for a sender will mostly be 0ns (or nominal
at best), providing an opportunity to reduce the number of exchanges required to send messages. 
Such a reduction has the potential to reduce the minimum end-to-end message latencies in \system, which are otherwise impacted by \rsv-\grt exchange overheads.

We therefore design an optimization that enables senders to send \textit{unsolicited} packet bursts immediately after sending a \rsv packet to the switch, without waiting for the \grt.  Unsolicited bursts are allowed only when the following two conditions are satisfied: (i) the timeslots at the input and output port known from a previous \grt are \textit{both} below a threshold $t$ 
and (ii), the information about the reservation is deemed to be recent, i.e., obtained within a certain interval of time (e.g., comparable to the time for a \rsv-\grt exchange). 
Condition (ii) is met when senders send consecutive bursts to the same destination. Packets that are sent unsolicited are marked by using a spare packet header field.
{
\small
\begin{algorithm}[t!]
	\caption{Scheduling at sender}
	\begin{algorithmic}[1]
	    \STATE \textbf{PARAMETERS:} $t$, $K$
		\STATE \textbf{INIT:}
		\STATE $lastChosenTimeslot \gets \{-1\}$
		\STATE $lastResponseTime \gets \{0\}$ 
		\vspace{4mm}
		\STATE \textbf{Function scheduleBurst}
		\STATE \textbf{INPUT:} packet burst with $K$ or fewer packets
		\STATE Send \rsv for burst
		\IF {$lastChosenTimeslot < t$ and $lastResponseTime$ is current}
		\STATE Send \textit{unsolicited} packet burst
		\ELSE 
		\STATE Call receive\grt
		\STATE Send packet burst after $waitingTime$
		\ENDIF
		\vspace{4mm}
		\STATE \textbf{Function receive\grt}
		\STATE \textbf{INPUT:} \grt
		\IF {$chosenTimeslot > t$} 
		    \IF {$lastChosenTimeslot < t$}
		    \STATE Resend packet burst corresponding to \grt
		    \ENDIF
		\ENDIF
		\STATE $lastChosenTimeslot \gets chosenTimeslot$
		\STATE $lastResponseTime \gets now()$ 
	\end{algorithmic}
	\label{algo:scheduler-sender}
	\end{algorithm}
}	


Algorithm~\ref{algo:scheduler-sender} shows the sender side scheduling logic. 
 As usual before sending a burst, \rsv packets are sent (line 7). However, if the
 \texttt{lastChosenTimeslot} was within \texttt{t} and the previous \grt was recent, an
 unsolicited packet burst is sent even before the next \grt arrives (line
 8-9). The switch schedule is modified to pass through unsolicited bursts,
 unless the reservation queues are larger than a threshold  $T>>t$, in which
 case unsolicited bursts are dropped (not shown in algorithm~\ref{algo:scheduler-switch}). 
 When
 the sender receives a \grt back, it determines whether the new
 \texttt{chosenTimeslot > t} (where \texttt{chosenTimeslot} is calculated using the equation from \Sec{sec:hostlogic}) and if so, resends the packet burst corresponding to the
 \grt (lines 16-20) at the right timeslot. This ensures that the packet burst
 arrives at the time it is expected at the switch, even if the unsolicited
 burst was dropped. However since \texttt{T} $>>$ \texttt{t}, it is also possible that both the
 unsolicited and scheduled packets arrive at the receiver, and the receiver
 has to deal with a small number of duplicate packets. 

In general, the threshold \texttt{t} is set to a sufficiently low value, so that the optimization 
only applies at very low loads. In our testbed implementation, we found \texttt{t} to be robust across a broad range of values and workloads. All our testbed experiments are run on a setting of \texttt{t=15}. Sweeping through \texttt{t=10} up to \texttt{t=25} does not show statistically significant changes to loss rates
or latency (of course, setting \texttt{t} to large values like 35 does lead to loss).

This optimization results in some transmissions arriving at the switch at the
same time under low load. We find that when there are only a few senders to a port, 
and there are continuous transmissions at a low rate, the optimization helps 
reduce the minimum message latencies; this is because the queuing caused by such 
simultaneous transmissions is smaller than the \rsv-\grt message exchange delay under low loads.



Tonic~\cite{tonic} and Sidler et.~al.~\cite{mothyfpga} have demonstrated that it is possible to place complex congestion control logic into NIC hardware. Since algorithm~\ref{algo:scheduler-sender} requires only a minimal subset of the supported logic (timed transmit), we believe it will be easier to implement in NIC hardware. Such an implementation will allow GPU and FPGAs (apart from CPUs) to access \system logic directly.



\subsection{Other design considerations}

\paragraph{Implications for intra-rack transports.}
We find that \system is able to provide congestion control to all traffic within the rack. One key advantage with TCP over \system is that transmissions with \system rely on current knowledge of network demand (using \rsv-\grt), rather than TCP's window estimate, which might become stale depending on the time of the last transmission within a flow. We are able to completely turn off TCP congestion control when using \system underneath in our 100G rack prototype(\secref{sec:impl}).



\paragraph{Handling failures.}
When a \system ToR fails, the entire rack fails, as is the case with non-\system rack. When a \system sender fails, its reservations on the ToR switch (up to $K$ outstanding for each connection) will need to be removed using external detection and recovery logic.


One way to detect a failed sender is to have \grt packets be additionally sent to receivers, and have receivers track pending transmissions. This adds a small overhead comparable to an ACK packet for every $K$ packets, and therefore trades-off a small amount of receiver bandwidth for better failure tolerance. When a sender crashes, or misbehaves, the receiver can detect the failure and reset the pending reservations at the switch. This scheme has an additional advantage of informing the receiver of upcoming transmissions; receivers can use this information to schedule the receiving process in time to receive the transmission to further drive down the end-to-end message delay with systems like~\cite{shenango}. We aim to look at this issue in the future and study its overheads.


\section{Implementation}
\label{sec:impl}
We have built a \Gbps{100} solution that uses \system in a rack. The switch scheduling algorithm (Alg~\ref{algo:scheduler-switch}) is implemented using a P4 program and runs on a ToR switch' programmable dataplane ASIC~\cite{tofino}. The switching delay with \system enabled is measured to be between \ns{346} (min) to \ns{508} (99.99\%ile), with a median delay of \ns{347} and standard deviation \ns{3}. \texttt{inReservation} and \texttt{outReservation}  are 32-bit register arrays updated in one switch pipeline stage; Since the switch has $64$ ports, each array consists of $64$ registers.

We use the Mellanox Messaging Accelerator (VMA)~\cite{mellanox-vma} to prototype the sender side scheduling support; we choose VMA instead of DPDK~\cite{dpdk} and the Linux network stack because VMA provides lower latencies on Mellanox NICs. The TCP/IP library integrated in VMA allows us to compare TCP and UDP performance with and without \system. We are able to turn off congestion control support in TCP when using \system underneath.
We also augment RDMA \reth with \system scheduling to mimic \system support for traffic generated by accelerators that might lack congestion control.  

\paragraph{\textbf{\system prototype topology and configuration}}
\system hardware prototype is a rack with $6$ servers, each equipped with a Mellanox ConnectX-5 \Gbps{100} NIC. Each server has two $28$-core Intel Xeon Gold 5120 \GHz{2.20} CPUs, \GiB{196} of memory, and runs Ubuntu 18.04 with Linux Kernel version 4.15 and Mellanox OFED version 4.4. The servers connect to a 6.5~Tbps programmable dataplane switch~\cite{tofino}, with 64 physical \Gbps{100} ports. The switch runs OpenNetworkLinux~\cite{openlinux} with Kernel version 3.16. The network MTU is 1500 bytes.

Our servers also connect to a Mellanox SN2700 switch using a second port on the Mellanox ConnectX-5 NIC. The servers synchronize over this out-of-band network using IEEE 1588 Precision Time Protocol (PTP)~\cite{ptp} using hardware NIC timestamps and boundary clock function at the Mellanox switch. 
Note, the time synchronization setup is only used for our
experiment infrastructure and not required to run \system. 

\paragraph{\textbf{\rsv-\grt exchange delay}}
We implement \rsv and \grt packets as 64-byte Ethernet control packets. \system continuously measures \rsv-\grt delays using hardware and software timestamping and uses these measures to correctly estimate packet transmission times (equation~\ref{eq:waittime}). We find that \rsv-\grt exchanges can have variable delay even in an unloaded network; the exchanges take between \us{1} (min), \us{1.06} (median) 
to 14us (max) NIC-to-NIC, using hardware timestamping and \us{1.98} (min), 2.05 (median) and \us{22.25} (max) using software timestamping in a network in an idle network.
Interestingly we see similar variance and tails when we transmit data close to line rate using \system. We anticipate that the performance that \system offers will get better if this variability is remedied. 
 

\paragraph{\textbf{Interfacing with application \texttt{send}}}
\system is prototyped on a user-space stack, where application \texttt{send} and \texttt{receive} is implemented in the same thread (as opposed to \texttt{send} and \texttt{receive} being handled in separate threads). We intercept function calls within the VMA library~\cite{mellanox-vma} that executes the actual \texttt{send} (\texttt{send\_lwip\_buffer} and \texttt{send\_ring\_buffer}) and send \rsv packets for every $K$ or fewer packets. Because the receive executes in the same thread, we also wait for the \grt before sending each packet burst; i.e., the application \texttt{send} call blocks until transmission completes. When employing the low-load optimization discussed in \secref{subsec:low_load}, even though we send packet bursts together with the \rsv, we wait to receive the \grt before transmitting the next burst. As such, we have not been able to eliminate the overheads due to \rsv-\grt exchange to the degree that the optimization design permits in our current implementation. A more favorable implementation of \system would execute \rsv-\grt receives in a separate thread or offload \rsv-\grt exchanges into the NIC. 
\section{Evaluation}
\label{sec:eval}
We evaluate \system \Gbps{100} prototype for the desired properties of losslessness, low latency and high utilization across various transports. 




\subsection{Experiment setup and methodology}
We use burst size $K=4$ and threshold $t=15$ in all experiments, configured as described in \S\ref{sec:design-k} and \S\ref{subsec:low_load}. We next describe the applications, transports, loads and traffic patterns we have evaluated.



\paragraph{\texttt{memcached}~\cite{memcached}} We evaluate the impact of worst-case background loads on latencies and throughput of single \texttt{memcached} client-server instances that reside on separate hosts. The key are 64B with \KiB{1} and \KiB{4} values. \texttt{memcached} uses TCP communication. The client executes reads (GET) continuously and the responses  compete with background traffic. We use reads rather than a mix of reads and writes because reads stress both the forward and reverse paths. Without background traffic,\texttt{memcached} introduces 2-\Gbps{3} network load.

\paragraph{Machine learning with vgg16~\cite{vgg16}} VGG16 is a popular convolutional neural network model used for large-scale image recognition. We emulate the network communication for VGG16 training, where gradients from each neural network layer are transferred over the network. Each parameter server in our set up receives messages from 4 workers at a total load of 70-\Gbps{89}.



\paragraph{Workload traces} We evaluate \system with traces generated from message-size distributions collected from various real-world modern application environments (W1-W5)~\cite{w1,w2,w4,w5} also used in Homa~\cite{homa}.  W1 is generated from a collection of memcached servers at Facebook, W2 from
a search application at Google, W3 from aggregated RPC-workloads 
from a Google datacenter, W4 is from a Hadoop cluster at Facebook and W5 is a web search workload used for DCTCP. The traces are generated in an open-loop with Poisson arrivals. These workloads represent modern database and analytics workloads that we expect rack-scale networks to run. 
The workload generator that replays these traces does not prioritize messages from one thread over the other. We believe this accurately mimics several user-space applications competing for the network independently across several cores.



\paragraph{Transport protocols} We evaluate the performance of 
\reth, UDP and VMA TCP over \system.  \reth and UDP represent accelerator transports that do not have end-host-based congestion control support. When we use \system underneath VMA TCP we turn off VMA's congestion control support\footnote{VMA only provides Cubic and Reno congestion control options, of which we have chosen Cubic}. We compare \system congestion control with VMA TCP Cubic and a receiver-driven congestion control scheme (RDS) based on Homa.

Our RDS implementation achieves close to \Gbps{100} line-rates by separating receiver-driven scheduling from data-processing; we schedule up to 4 packets when possible and use a lock-free mechanism for communication between the scheduling and data-processing threads. When packets are lost, the receiver can detect these losses by monitoring out-of-order packet arrivals. In these cases, the receiver notifies the sender of the lost packets immediately (rather than waiting for a timeout~\cite{homa}). This helps reduce delays due to a majority of lost packets. Sometimes an entire burst of packets is lost, and in these cases, the receiver cannot effectively determine if the packets are lost or delayed. Instead they timeout after \mis{1} and send a lost-packet notification to the sender.
Senders send up to \KiB{61}, the bandwidth-delay product in our system blindly as fast as they can (similar to RTTbytes in~\cite{homa}). If these initial packets are lost, the sender times out after \mis{1}, and resends the packets again.

Server-server \reth provides an optimal baseline for comparison with \system. \reth is unhindered by congestion control, and transmits messages as fast as they arrive. No congestion control scheme can achieve smaller delays than \reth. Recently \cite{ndp, homa} showed that RDS-like mechanisms can provide significantly smaller latencies than available congestion control schemes. Therefore we choose to implement and compare \system with RDS on the testbed. We also compare \system with VMA TCP Cubic, because VMA touts impressive latency improvements.

We are unable to make a fair-comparison with DCTCP~\cite{dctcp} and other available congestion control schemes in the linux kernel because \system is only implemented at user-space (using VMA), and does not experience the overheads of the linux kernel.

\paragraph{Traffic patterns, link loads and latency evaluation}
We evaluate \system under incast, outcast, and shuffle traffic patterns. Incast helps demonstrate the lossless behavior of \system. Outcast is \system's worst-case traffic pattern, because it stresses senders that not only transfer data but also have to do \rsv-\grt signalling.



 We control the network load at each server by injecting traffic using multiple threads pinned to different cores. We experiment with both persistent congestion caused by multiple long running background flows, and with microbursts caused by replaying W1-W5 traces from multiple threads.


We present results of 99\%ile and  median latencies measured in user-space; for memcached, we measure two-way request-response delays;
for other workloads, we measure one-way latencies, i.e., the time from message arrival at the sending thread to the time when the message is delivered at the receiving thread. 


The rest of the evaluation section summarizes our findings.

\begin{figure}[!ht]
\vspace{-10pt}
  \begin{minipage}{0.40\textwidth}
     \centering
	   \includegraphics[width=.95\linewidth]{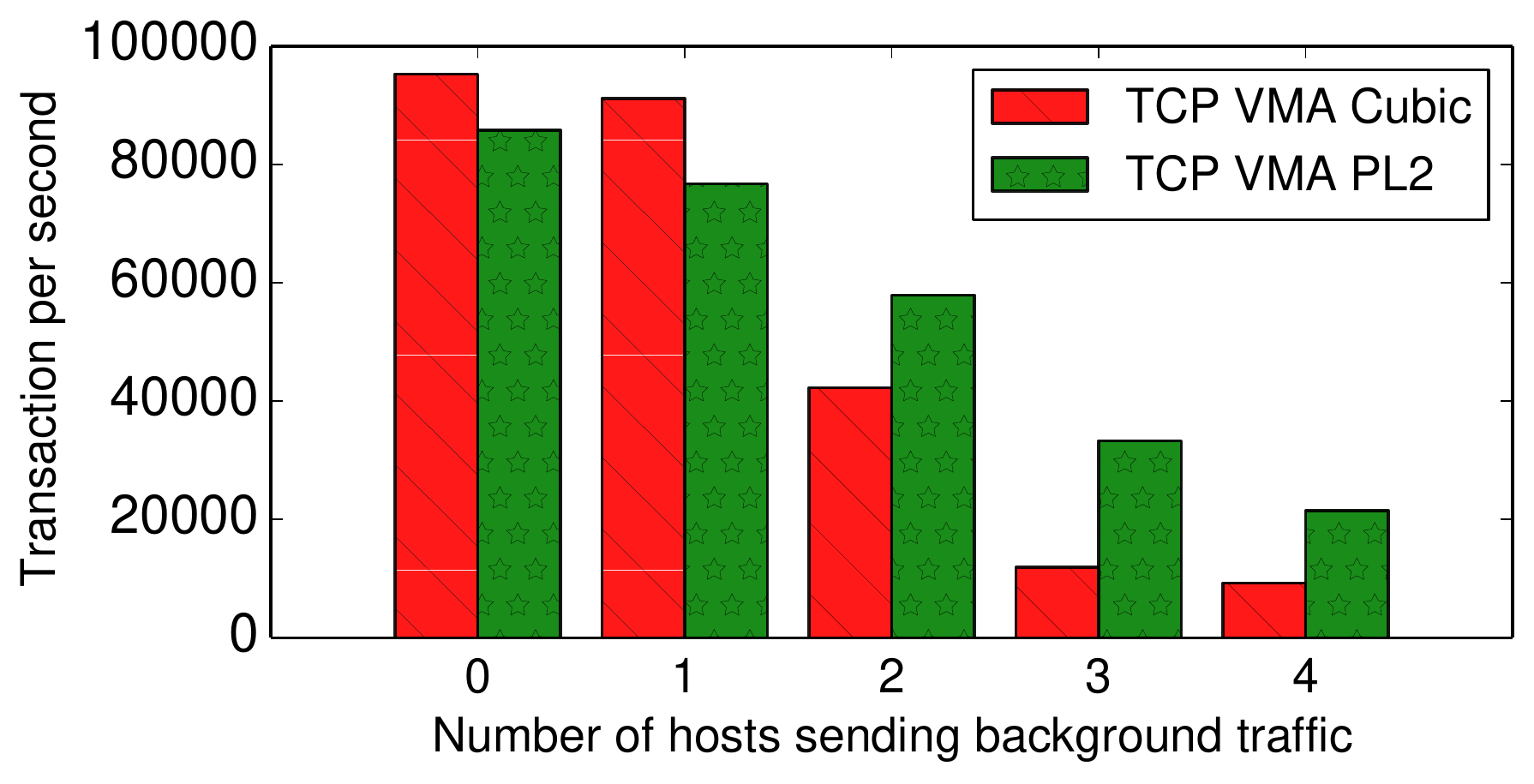}
	   \caption{\small{Memcached throughput with competing TCP traffic.}}
	   \label{fig:tcp-mem-tput}
  \end{minipage}
  \begin{minipage}{0.40\textwidth}
     \centering
	   \includegraphics[width=.95\linewidth]{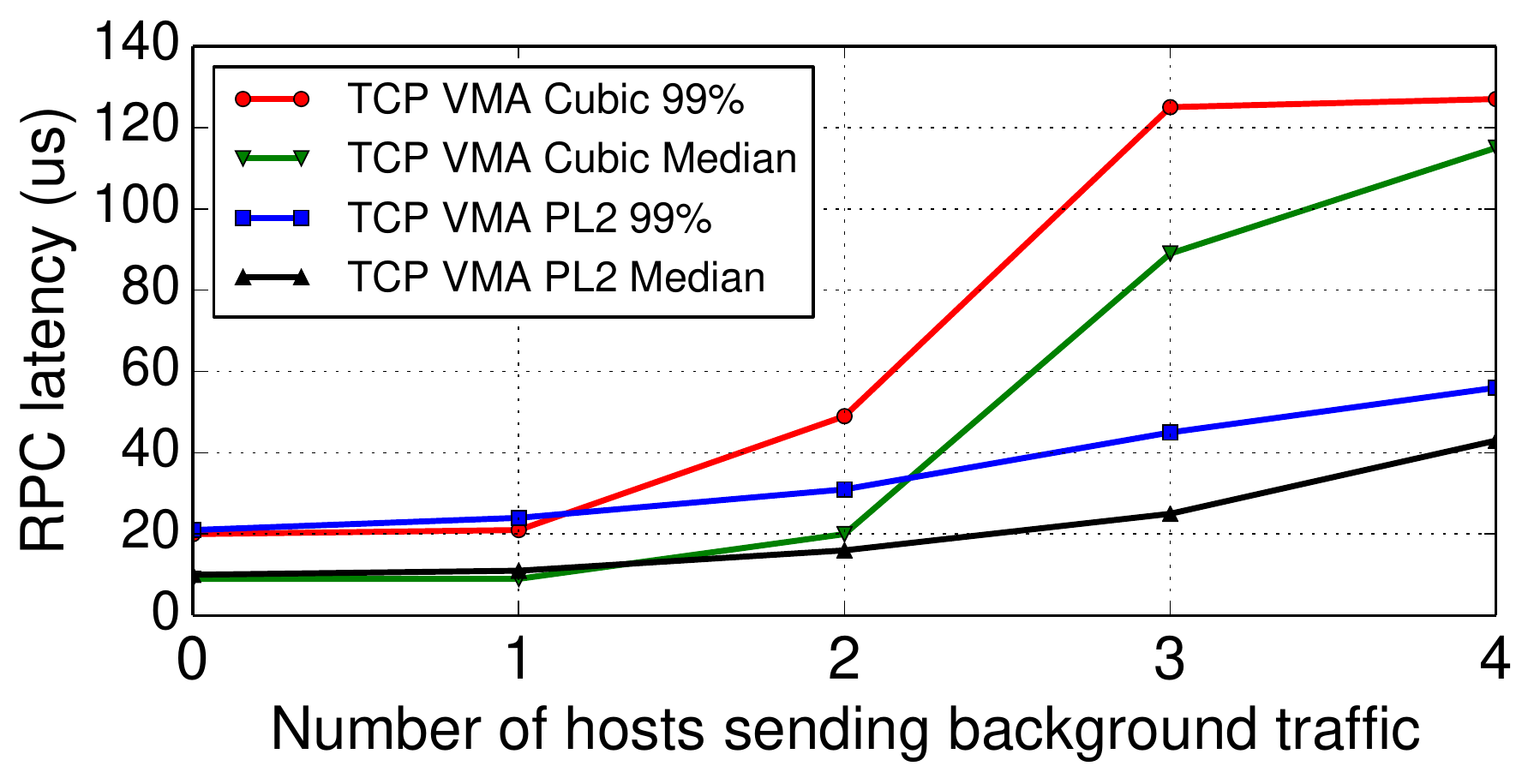}
	   \caption{\small{Memcached latencies with competing TCP traffic.}}
	   \label{fig:tcp-mem-lat}
  \end{minipage}
      \begin{minipage}{0.40\textwidth}
     \centering
	   \includegraphics[width=.95\linewidth]{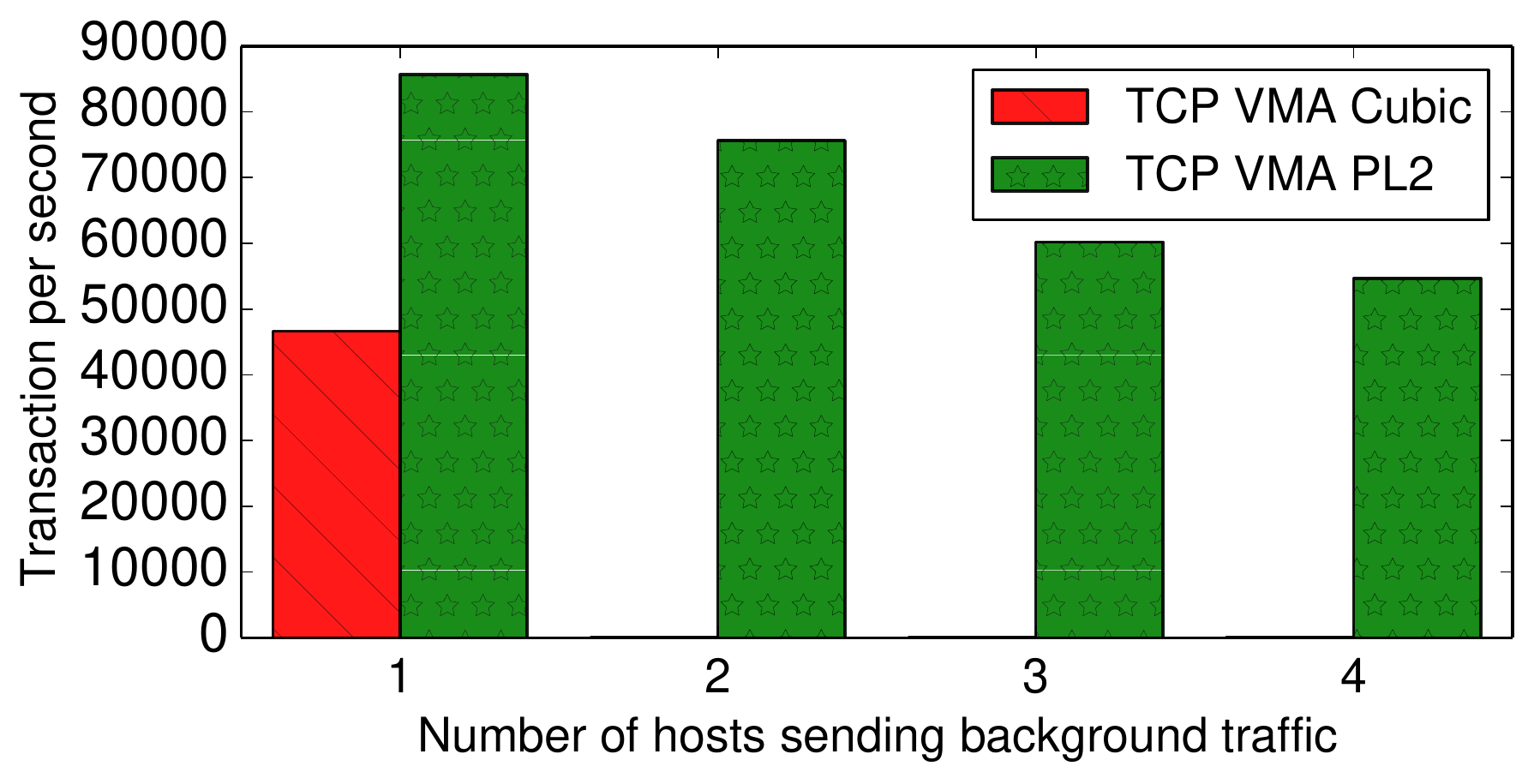}
	   \caption{\small{Memcached throughput with competing UDP traffic.}}
	   \label{fig:udp-mem-tput}
  \end{minipage}
      \begin{minipage}{0.40\textwidth}
     \centering
	   \includegraphics[width=.95\linewidth]{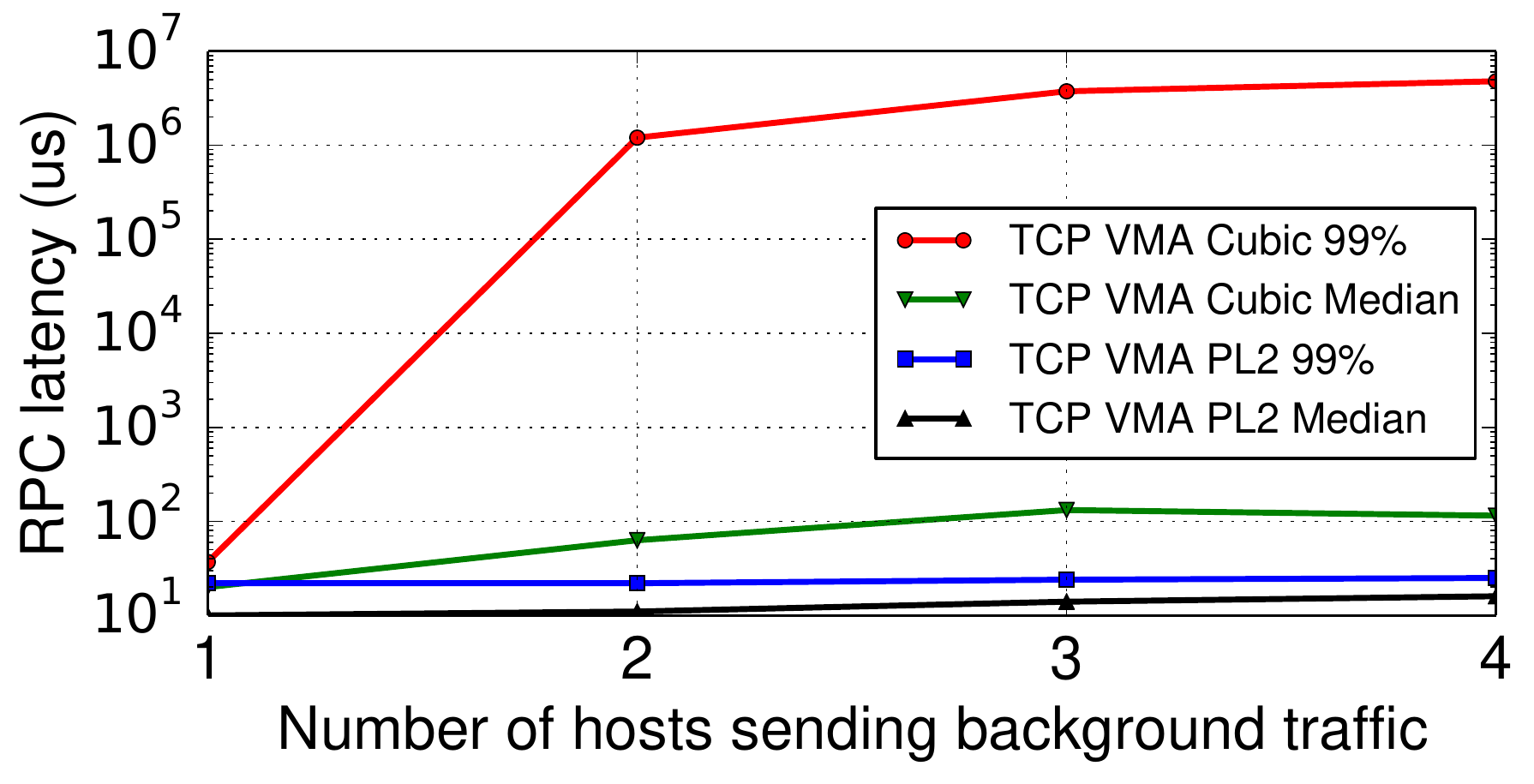}
	   \caption{\small{Memcached latencies with competing UDP traffic.}}
	   \label{fig:udp-mem-lat}
  \end{minipage}
\end{figure}

\begin{figure}
   \begin{minipage}{0.45\textwidth}
     \centering
	   \includegraphics[width=.95\linewidth]{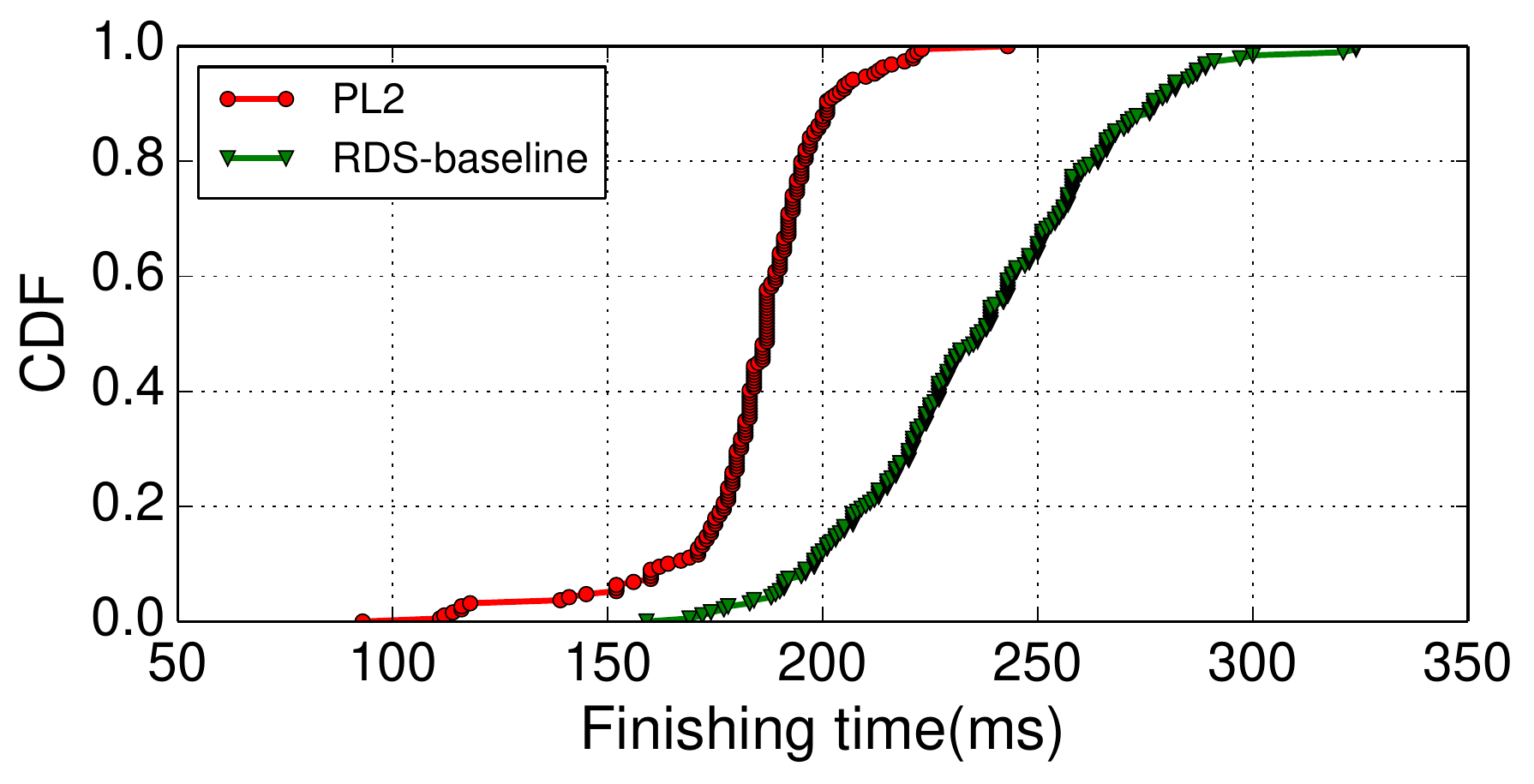}
	   \caption{\small{CDF of latency for all iterations in VGG16 model training using \system and RDS.}}
	   \label{fig:vgg16-finish}
   \end{minipage}
   \end{figure}

\subsection{End-to-end application performance}

We experiment with two real-world applications, \texttt{memcached} and vgg16 training. When \texttt{memcached} competes with heavy (incast) background traffic with congestion control support, \system can ensure up to 2.3x lower rpc-latencies, and 1.8x corresponding improvement in application throughput compared to VMA TCP Cubic. When it competes with traffic without congestion control(the kind of traffic we expect from accelerators), \texttt{memcached} over \system still sees similar latencies and throughput; whereas without \system, \texttt{memcached} rpcs fail to complete due to severe losses. This demonstrates that \system is able to provide fabric-level congestion control for all transports that use it.

Experiments with VGG-16 training show that \system can reduce network transfer times per iteration by 30\% compared to RDS (222ms vs 321ms 99\%ile latency per iteration); we achieve a 20\% speed up for 100 iterations (9.5s versus 12s). 

\subsubsection{\system keeps memcached 99\%ile latencies below \us{60} even with 400\% offered load}
\label{sec:eval-memcached}
Figures \ref{fig:tcp-mem-tput}-\ref{fig:tcp-mem-lat} show \texttt{memcached} transaction throughput and response latencies with TCP background traffic. The graphs shown are with \KiB{4} values; the results for the same experiment with \KiB{1} values show the same trend.
\figref{fig:tcp-mem-tput} shows how memcached throughput (in transactions per second) reduces with increasing TCP background traffic. The red bar shows throughputs with memcached over TCP over \system (TCP VMA \system), and the green bar shows throughputs with memcached over TCP with Cubic congestion control (TCP VMA Cubic).

When there is no background traffic (bars for 0 hosts sending background traffic), \system slows down memcached by a small amount(8\%), even though the RPC latencies are similar to TCP VMA Cubic (\figref{fig:tcp-mem-lat}). \system \rsv-\grt exchange overheads contribute to reduced throughput seen with TCP VMA \system. The same effect is seen in the case where the \texttt{memcached} host also sends background traffic that utilizes around 50\% of the downlink to the \texttt{memcached} client (\Gbps{50}). Once we add background traffic load from additional servers, the load on the receiving link increases to 80-90\% and the \texttt{memcached} latencies go up much faster for TCP VMA Cubic than TCP VMA \system as seen in \figref{fig:tcp-mem-lat}; consequently \texttt{memcached} has 0.5x lower throughput with TCP VMA Cubic.
This is despite the fact that the competing background traffic that runs using TCP VMA \system has 1-\Gbps{5} more load than the background traffic that runs on VMA TCP Cubic. 
The 99th-percentile RPC latency with VMA TCP Cubic is \us{127} as opposed to \us{56} with \system with 4-way TCP incast background traffic.

\subsubsection{\system keeps memcached latencies low~(\us{25}) even when competing with traffic with no end-host-based congestion control}
Figures \ref{fig:udp-mem-tput}-\ref{fig:udp-mem-lat} show \texttt{memcached} transaction throughput and response latencies with UDP background traffic (memcached traffic still uses TCP). Since UDP has no congestion control, it presents particularly severe competition to memcached traffic. Our goal with this experiment is to show how \system enables multiple transports with or without congestion control to co-exist within a rack.

\figref{fig:udp-mem-tput} shows the throughput of \texttt{memcached} as we increase UDP background load. Without \system, the \texttt{memcached} benchmark does not complete due to severe losses. 
\figref{fig:udp-mem-lat} shows the 99\%ile and median memcached RPC latencies with VMA TCP Cubic and \system. Memcached sees 5s 99th-percentile latency without \system (due to drops) as opposed to \us{25} with \system with 4-way UDP incast background traffic.


\begin{figure*}[!htb]
      \begin{subfigure}{0.33\textwidth}
     \centering
	   \includegraphics[width=.95\linewidth]{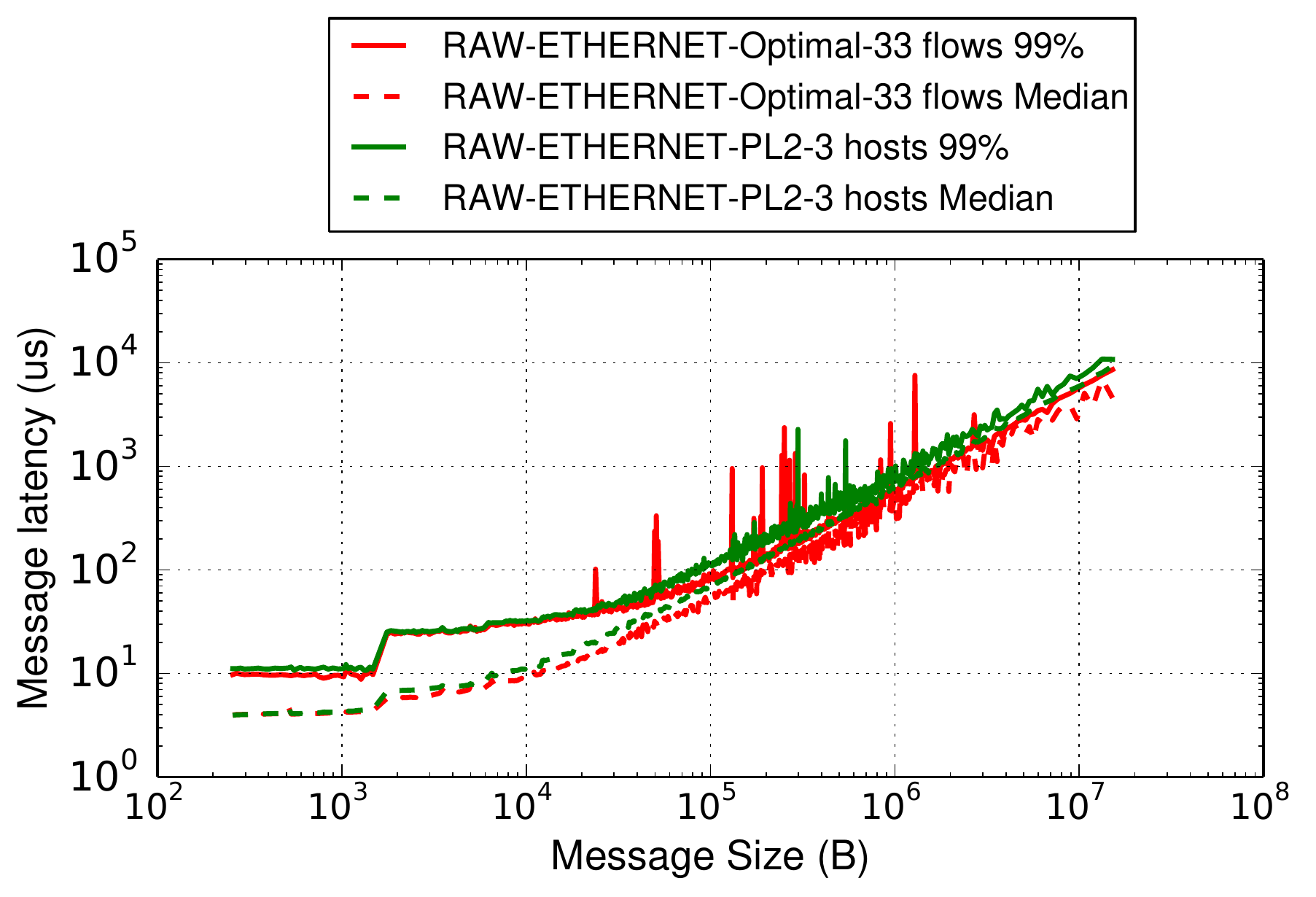}
	   \caption{\label{fig:raw-incast-w3}\small{W3 All RPC (\reth)}}
   \end{subfigure}\hfill
         \begin{subfigure}{0.33\textwidth}
     \centering
	   \includegraphics[width=.95\linewidth]{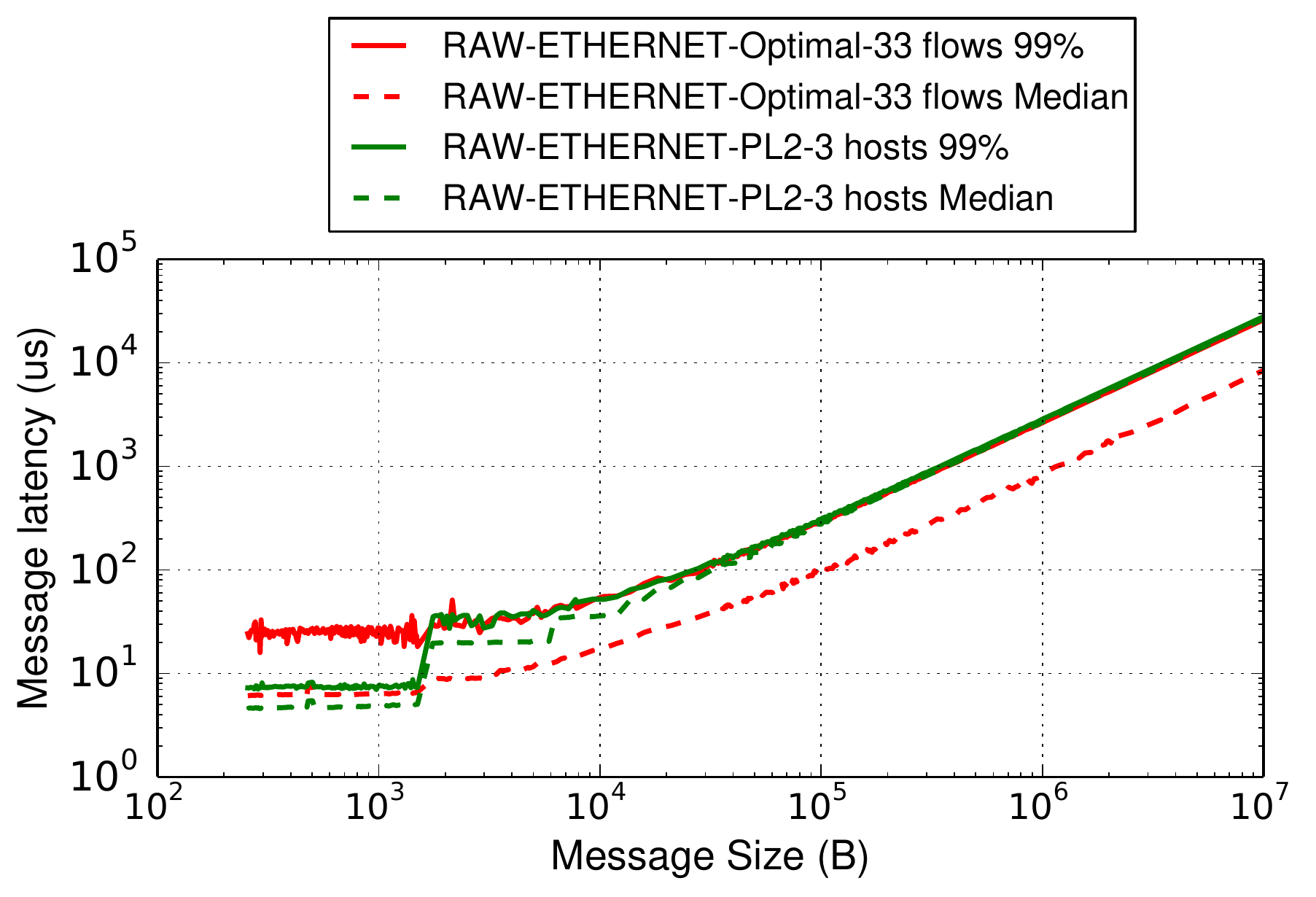}
	   \caption{\label{fig:raw-incast-w4} \small{W4 Hadoop (\reth)}}
   \end{subfigure}\hfill
      \begin{subfigure}{0.33\textwidth}
     \centering
	   \includegraphics[width=.95\linewidth]{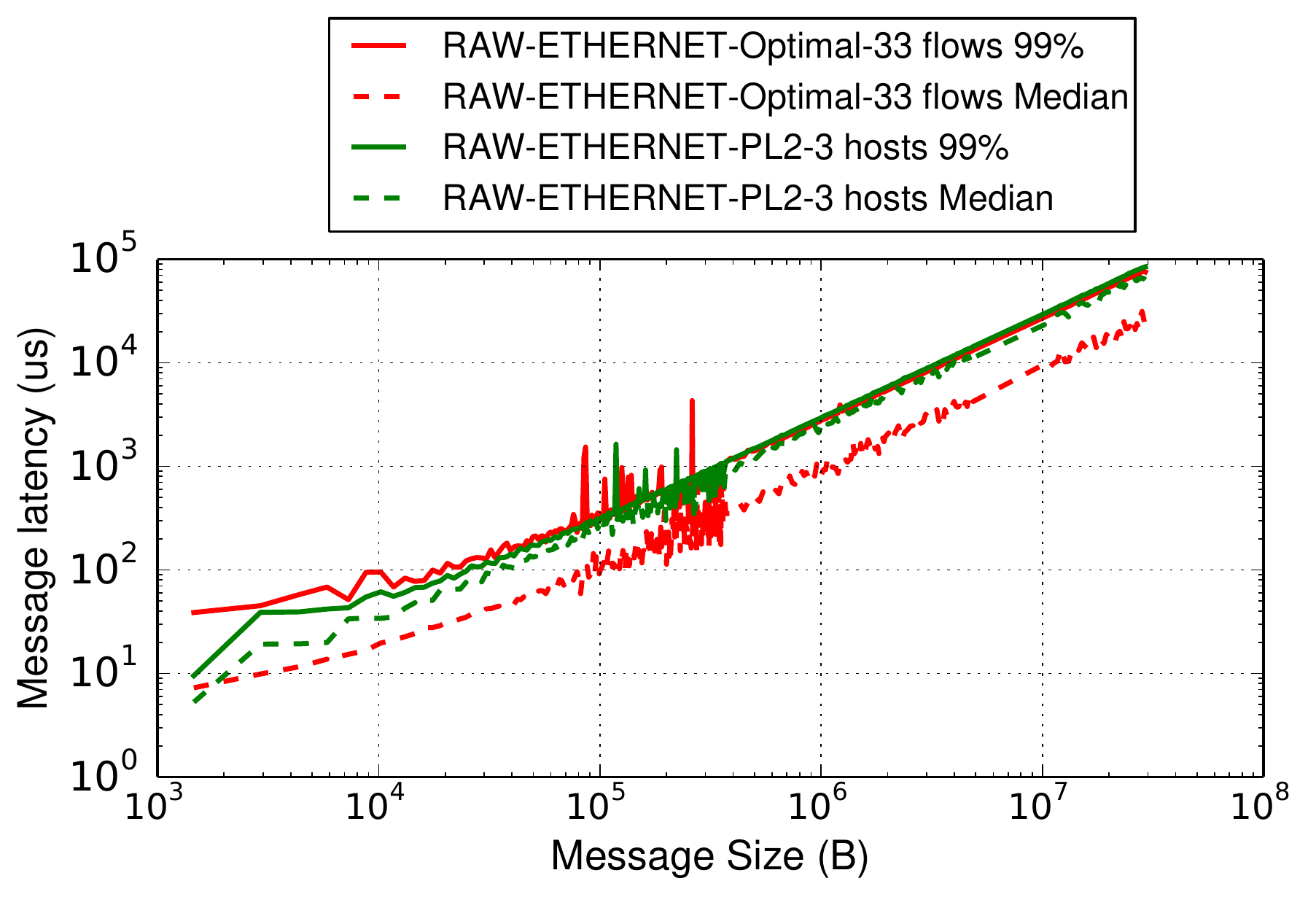}
	  \caption{\label{fig:raw-incast-w5} \small{W5 Search (\reth)}}
   \end{subfigure}\hfill
   \begin{subfigure}{0.33\textwidth}
     \centering
	   \includegraphics[width=.95\linewidth]{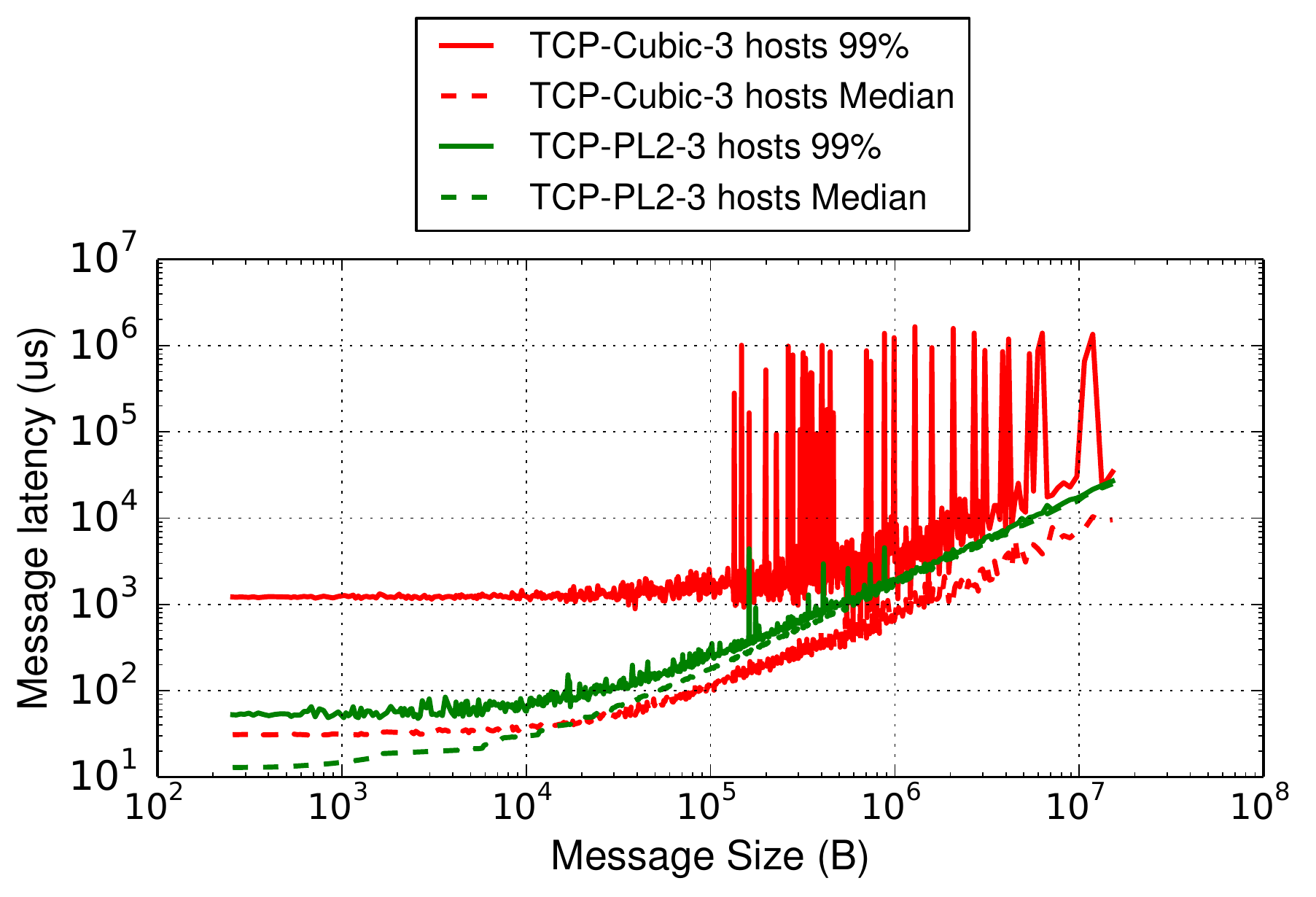}
	   \caption{\label{fig:tcp-w3} \small{W3 All RPC (TCP)}}
   \end{subfigure}\hfill
    \begin{subfigure}{0.33\textwidth}
     \centering
	   \includegraphics[width=.95\linewidth]{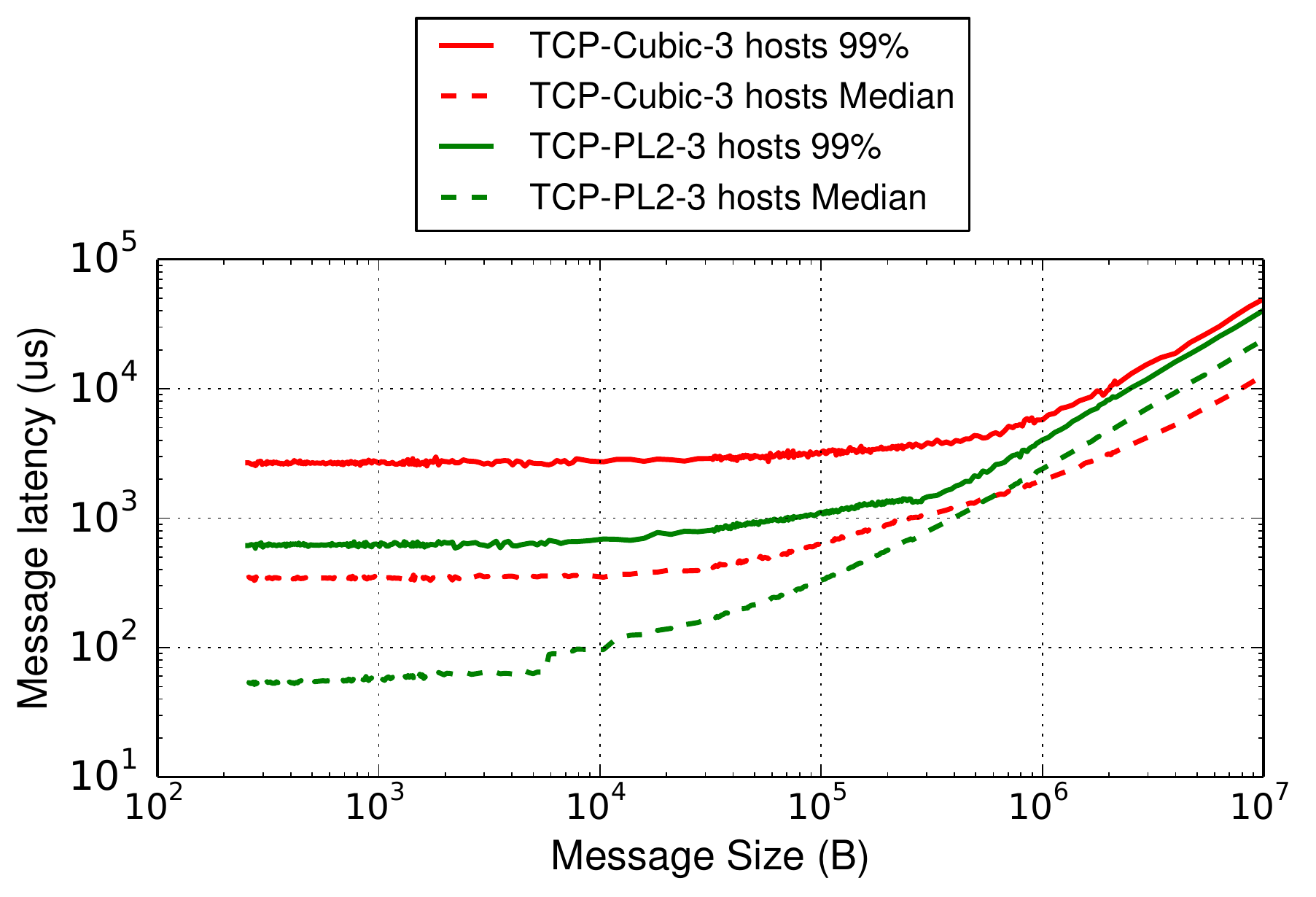}
	   \caption{\label{fig:tcp-w4}W4 Hadoop (TCP)}
   \end{subfigure}\hfill
   \begin{subfigure}{0.33\textwidth}
     \centering
	   \includegraphics[width=.95\linewidth]{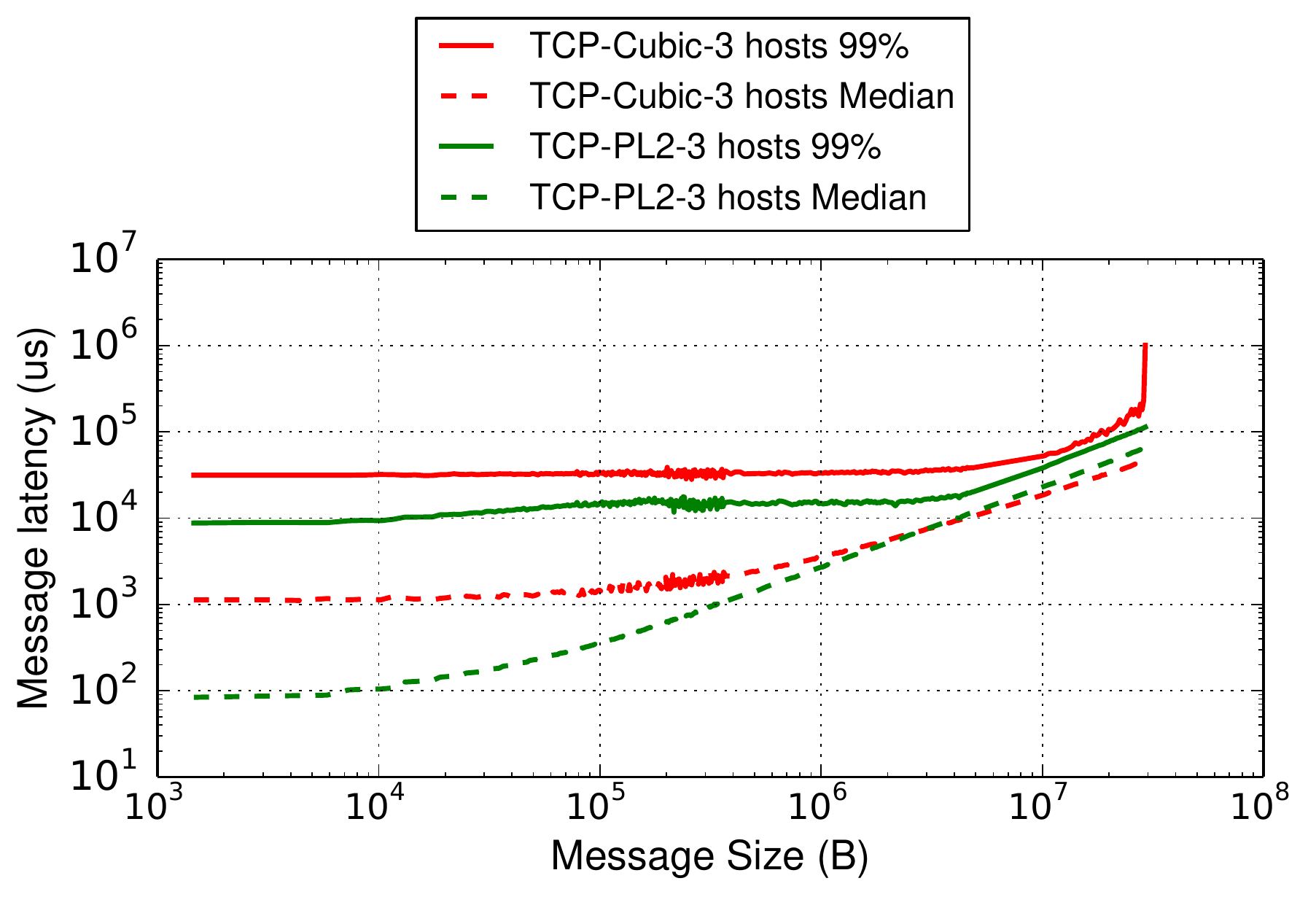}
	   \caption{\label{fig:tcp-w5}\small{W5 Search (TCP)}}
   \end{subfigure}\hfill
     \caption{Message latencies of workloads W3-W5 for \reth and TCP}
\end{figure*}

\begin{figure*}[!t]
   \begin{subfigure}{0.33\textwidth}
     \centering
	   \includegraphics[width=.95\linewidth]{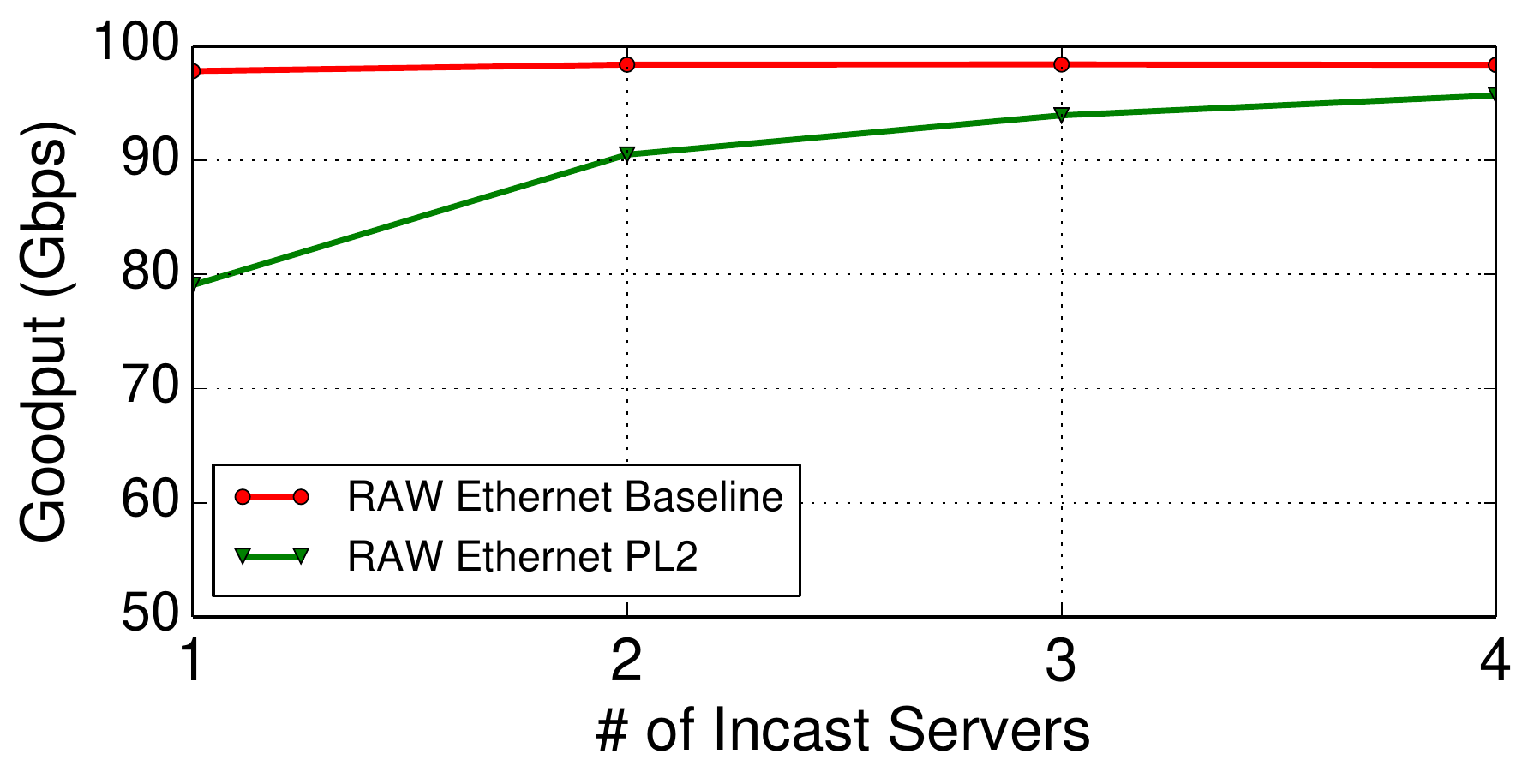}
	   \caption{\label{fig:raw-incast-tput} \footnotesize{Throughput over Ethernet}}
   \end{subfigure}\hfill
   \begin{subfigure}{0.33\textwidth}
     \centering
     \includegraphics[width=.95\linewidth]{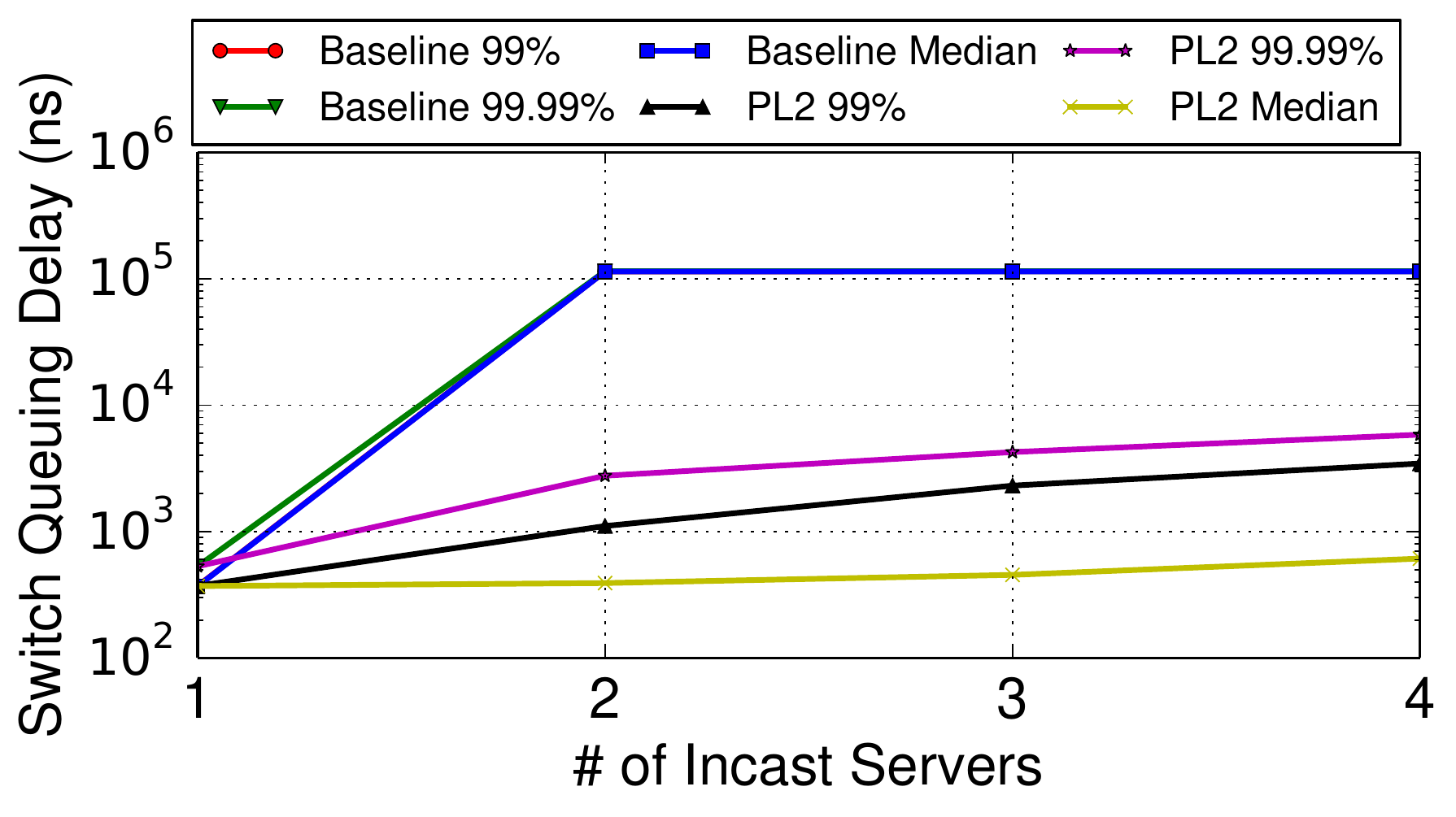}\caption{\label{fig:raw-incast-queue-length} 
     \footnotesize{Switch queuing delays for \reth and \system} }
   \end{subfigure}\hfill
   \begin{subfigure}{0.33\textwidth}
     \centering
     \includegraphics[width=.95\linewidth]{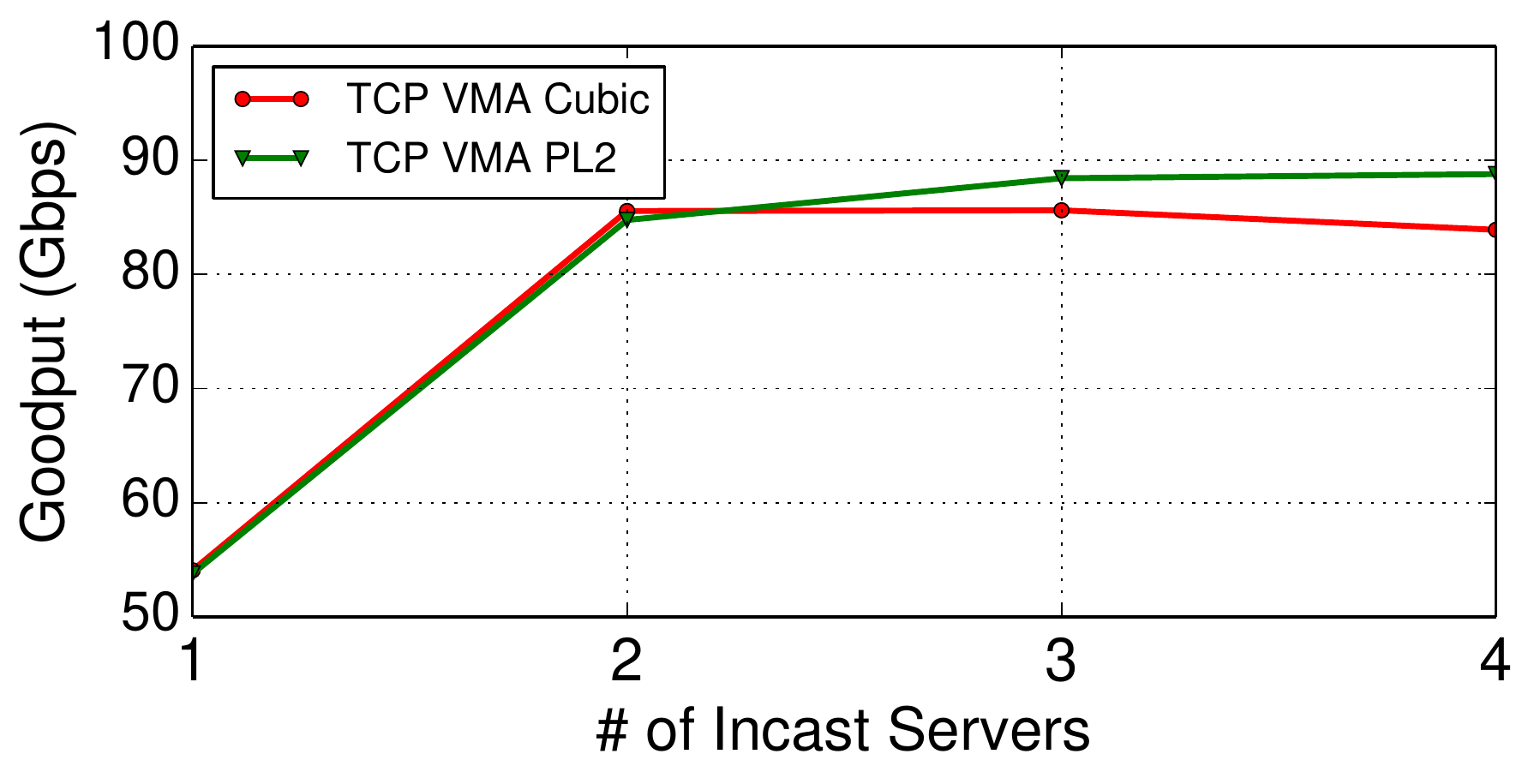}
	   \caption{\label{fig:tcp-incast-tput}\footnotesize{Throughput over TCP}}
   \end{subfigure}\hfill
   \caption{Incast evaluation comparing against \reth and TCP.}
\end{figure*}

Once we introduce background traffic from 2 or more hosts, we find that memcached incurs severe losses without \system, and its throughput drops to 3-7 transactions per second, while the tail latency shoots up to several seconds (\figref{fig:udp-mem-lat}). However, with \system, the memcached RPC latencies do not degrade with such severity (even the tail latencies remain steady, with some increase in the median), and memcached continues to have throughputs similar to our experiment with TCP background traffic. This is because \system keeps the UDP background traffic from 2-4 hosts within 75-\Gbps{90}. 

\subsubsection{\system improves training latencies for vgg16 by 30\%}
The communication pattern for exchanging gradient updates in VGG16 is inherently a shuffle process. We have $4$ hosts sending data as workers and $2$ hosts as parameter servers, which receives gradients from all the workers. The gradient data (\MiB{500}) is partitioned according to VGG16 architecture. The aggregate incoming rate to each parameter server is less than the line rate to ensure the network is not a bottleneck. We repeat this process  $100x$ and measure the finishing time of transferring the entire gradient set at each iteration. 

Figure~\ref{fig:vgg16-finish} shows the CDF of iteration times using the receiver-driven scheme and \system. We did not prioritize small messages with the receiver-driven scheme like Homa~\cite{homa} because it interferes with the ordering of messages that the training set expects. As such, we find that the $99$th percentile finishing time of each iteration in the receiver-driven scheme is $1.45x$ than that of \system. 
To finish
$100$ gradients sets from each worker, \system takes 9.5s in total while the receiver-driven scheme takes 12s.
The receiver-driven approach is slower because the receiver is unaware of traffic from the sender to other hosts: it frequently allocates time slots when the sender is already sending to another receiver. This leads to lower link utilization. 
\system does not have this issue because \system counts the queuing both at the sender and receiver.

\subsection{W1-W5: Near-optimal p99 latencies}
\label{sec:eval-w1to5}

We compare how close message latencies for W1-W5 are to optimal with 3-way incast using \reth over \system. We do this by using a baseline where W1-W5 are run using just \reth between two servers. Since the baseline encounters no contention, and \reth transmits messages as soon as they arrive (with no congestion control), the message latencies it encounters are close to optimal. We call this scheme \oreth. When comparing 3-way incast results with \reth over \system to \oreth, we try to keep the network load equivalent. We do so by using the same number of threads in total to run the workload across the two schemes (11 threads per server, 33 threads in total).

\begin{figure}[hb]
 \centering
   \begin{minipage}{0.40\textwidth}
     \centering
	   \includegraphics[width=0.8\linewidth]{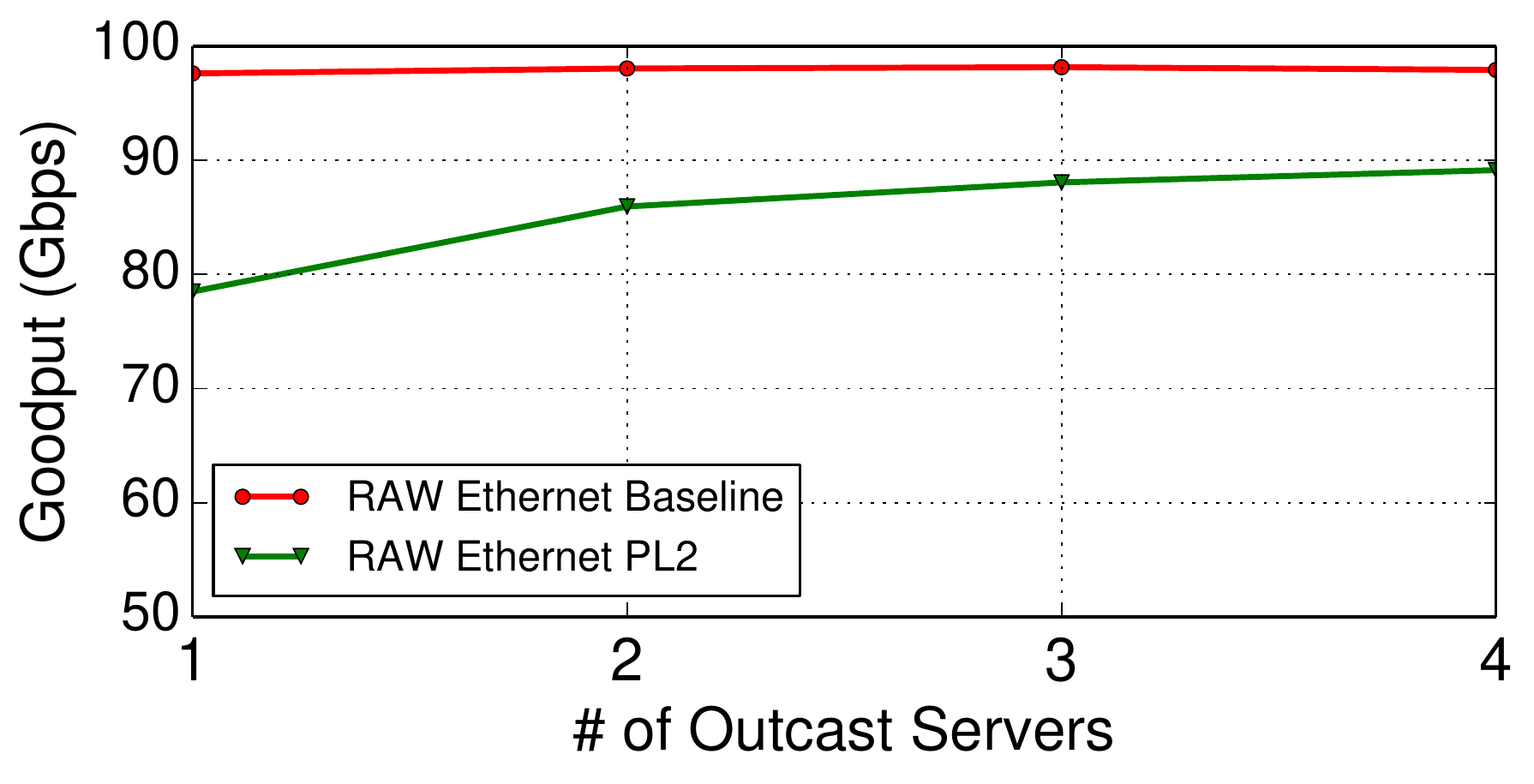}
	   \vspace{-4mm}
	   \caption{\label{fig:raw-outcast-tput}\footnotesize{Outcast throughput comparison \reth vs. \pleth}}
   \end{minipage}\hfill
\end{figure}

Figs. \ref{fig:raw-incast-w3}-\ref{fig:raw-incast-w5} present the message latencies for workload traces W3-W5 over \reth. For these workloads, 3-way incast achieves 40-\Gbps{90} throughput.  \pleth has 99\%ile latencies close to \oreth, and the throughputs obtained with and without \system are similar (except in the case of \figref{fig:raw-incast-w4} where \pleth gets slightly lower throughput compared to \oreth for W4 (\Gbps{86} vs. \Gbps{89}) due to \rsv-\grt exchange overhead). The impact of \rsv-\grt exchanges are more prominent in the median latencies in Figs. \ref{fig:raw-incast-w4} and \ref{fig:raw-incast-w5}. 

We see latency spikes in Figures \ref{fig:raw-incast-w3} and \ref{fig:raw-incast-w5}; these are outliers, when the message arrivals exceeded the capacity of our workload generator 
when it ran out of available worker threads to transmit the next message.

\subsubsection{W1-W5 over TCP: 10x lower 99\%ile latencies}
We also compare 3-way incast of W1-W5 with TCP-over-\system to 3-way incast with TCP Cubic.
Figures \ref{fig:tcp-w3}, \ref{fig:tcp-w4}, \ref{fig:tcp-w5} show that TCP-Cubic has worse median and 99\%ile latencies than TCP over \system, except in the case of \figref{fig:tcp-w3}. VMA TCP Cubic has better latencies with W3, because it undergoes a congestion collapse and achieves only \Gbps{43}, as compared to \Gbps{81} with TCP over \system. In all other cases, we find that VMA TCP Cubic achieves throughputs close to TCP with \system while having retransmissions due to message loss (TCP with \system has no losses). We believe these graphs show the stable-queuing effect of proactive congestion control.

Our workload generator is unable to generate greater than \Gbps{5} traffic for W1 and W2 even in an incast scenario because the master thread cannot keep up with assigning messages to connections within the inter-arrival times. We find that when workloads impose such light loads, \system does not give significant benefits over VMA TCP Cubic, nor does it cause significant degradation; we omit these results for space and refer the reader to our results with memcached with no background traffic in \figref{fig:tcp-mem-tput}.

\subsection{Throughput implications}
\figref{fig:raw-incast-tput} shows the aggregate throughput achieved by  \pleth in comparison to \breth, with incast, while not having loss. The x-axis shows the number of hosts that are sending out persistent traffic to the same receiver and the y-axis shows the throughput in Gbps. Each host starts 12 flows (pinned to separate cores on the sender), and sends 24 million \KiB{6} messages (4 MTU sized packets).
\breth achieves line rate for the case of one host sending traffic to another host without any 
packets drop. However, as the number of sending hosts $n$ increases, we find that only  
only $1/n$ packets are delivered to the receiver (all senders are able to send at line-rate).  \pleth has no losses but caps server-to-server throughput to \Gbps{80}. The throughput depends on the number of flows in \system (12), because each flow has at most one outstanding \rsv when the network is loaded.

\figref{fig:raw-incast-queue-length} shows the switch queuing delays for the same experiment. 
As can be seen \breth has uncontrolled queue lengths, whereas \pleth achieves stable queueing with low variance. 

\figref{fig:tcp-incast-tput} shows \system's aggregate throughput in comparison to TCP cubic, with incast traffic. The experiment settings mimic the \reth experiment. TCP cubic experiences 233, 313, 404 retransmissions for 2, 3, and 4 host incasts. In the case of congestion controlled TCP traffic, traffic over \system sees the same (or higher) throughput as the baseline, i.e., \system's proactive scheduling is comparable to TCP's reactive scheduling, while preventing losses.

Outcast traffic pattern stresses \system senders to the maximum extent. With \reth outcast, as shown in \figref{fig:raw-outcast-tput} (with the same settings as above),  the maximum throughput \system can achieve is capped at \Gbps{90} (as opposed to \Gbps{97} in the case of incast). This drop (10\%) is due to the overhead of \rsv-\grt exchanges at the sender; it is the price \system pays for it's proactive scheduling design.




\section{Limitations}
\label{sec:disc}
\paragraph{\textbf{Utilization}}
\system cannot fully utilize the Ethernet capacity available, although gets close (up to \Gbps{96} utilization with incast, and up to \Gbps{90} with outcast). \rsv and \grt signalling overheads take up some spare capacity; in our implementation \system adds a minimum of 2\% bandwidth overhead (128B of overhead for $K=4$ MTU~(1500B) packets). In the worst case when all network packets are at 64B, the overheads are comparable to the demand in the network.   However, this is not the common-case for which \system is designed. 
Variation in hardware delays due to DMA transfer latencies can introduce scheduling inefficiencies. \system also leans towards preventing losses in selecting the time duration to wait before sending packets, and thus can leave some capacity unused.

\paragraph{\textbf{Fairness}}
Like Homa~\cite{homa}, \system is unfair but not to large messages. \system is unfair because the switch scheduler schedules bursts in FIFO order of receiving \rsv packets. The scheduling size $K$ limits this unfairness; under loaded conditions, until the scheduled $K$ packets are sent, the next \rsv packet is not sent. We believe that \system scheduler design should change to eliminate this unfairness once switch-hardware is able to handle richer reservation logic.
\paragraph{\textbf{In-network priorities}}
\system does not implement in-network priorities; this design also stems from the limitations of switch-hardware processing \rsv packets in FIFO order. 
However, given that \system does not drop packets, and ensures that at any point only a few ($K$) packets from different senders are in flight to a receiver, we anticipate that in-network priorities will not have a big role to play in further reducing latencies with \system.
Rather the order in which \rsv packets are admitted into the network would play a bigger role, which in turn is determined by (i) external policies that govern rate at which \rsv packets can be sent by an application;  (ii) process scheduling prioritization within the rack; for e.g., how many cores the communicating processes are given, how often applications are scheduled in comparison to others; and (iii) application's internal logic. We aim to study these implications in our future work.
\paragraph{\textbf{Handling inter-rack traffic}}
 To the \system scheduler, traffic leaving the rack using a port on the ToR is no different from intra-rack traffic; the traffic exiting will also be scheduled using the timeslot reservation scheme. However this traffic will not be co-ordinated with other inter-rack traffic destined to an external rack.
 With \system's current design, static bandwidth reservations will be required to ensure that traffic entering a \system rack will not disrupt \system guarantees for intra-rack traffic, or be dropped altogether; e.g., 10\% of rack bandwidth is reserved for ingress traffic. \system can reflect such reservations either in the length of timeslot or the reservation increments. 
\section{Conclusion}
\label{sec:conc}
In this paper, we present the \system rack-scale network architecture designed to convert Ethernet into a reliable, predictable latency, high-speed interconnect for high density racks with accelerators by leveraging new capabilities in programmable switches.  Our hardware prototype demonstrates that \system does so by providing practical and tightly coordinated congestion control. Further, we achieve our goals without any knowledge of workload characteristics or any assumptions about future hardware stacks. 
We believe that the design presented in this paper will spur new ideas around switch and NIC hardware, how they interface with and simplify the network and applications stack. 

This work does not raise any ethical issues.

\bibliographystyle{plain}

\bibliography{reference1}

\end{document}